\documentclass[journal]{new-aiaa}

\usepackage{graphicx}
\usepackage{amsmath}
\usepackage[font=small]{caption}
\usepackage{xcolor}

\usepackage{subcaption}

\usepackage{longtable,tabularx}

\usepackage[toc,page]{appendix}

\newcommand{\minus}{\scalebox{0.75}[1.0]{$-$}}
\renewcommand{\thesubsection}{\Roman{section}.\Alph{subsection}}

\title{Closed-loop control of a piezo-fluidic amplifier}

\author{Chris Nicholls\footnote{PhD Student, Oxford Thermo-fluids Institute, University of Oxford, christopher.nicholls@eng.ox.ac.uk.}$\left.\right.$ and Marko Bacic\footnote{University Research Lecturer, Oxford Thermo-fluids Institute, University of Oxford. Member AIAA. On secondment from Rolls-Royce Plc.}}
\affil{Oxford Thermo-fluids Institute, University of Oxford, Oxford, OX2 0ES, United Kingdom}

\begin{document}

\maketitle
\thispagestyle{empty}
\pagestyle{empty}

\singlespacing
\begin{abstract}
Fluidic valves based on the Coand\u{a} effect  are increasingly being considered for use in aerodynamic flow control applications. A limiting factor is their variation in switching time, which often precludes their use. The purpose of this paper is to demonstrate the closed-loop control of a recently developed, novel piezo-fluidic valve that reduces response time uncertainty at the expense of operating bandwidth. Use is made of the fact that a fluidic jet responds to a piezo tone by deflecting away from its steady state position. A control signal used to vary this deflection is amplitude modulated onto the piezo tone. Using only a pressure measurement from one of the device output channels, an output-based LQG regulator was designed to follow a desired reference deflection, achieving control of a 90 $\text{ms}^{{\minus}1}$ jet. Finally, the controller's performance in terms of disturbance rejection and response time predictability is demonstrated.
\end{abstract}
\section*{Nomenclature}
\begin{minipage}[t]{0.49\textwidth}
{\renewcommand{\arraystretch}{1.0}
\noindent\begin{tabular}{lp{7cm}}
	
\multicolumn{2}{l}{\textbf{Miscellaneous}}\\
$\alpha$ & Feed-forward gain (-)\\
${\delta}\mathbf{x}(t)$ & State deviation from equilibrium (-)\\
$\hat{{\delta}\mathbf{x}}(t)$ & State deviation estimate (-)\\
${\delta}y(t)$ & Output deviations from reference (-)\\
${\delta}y'(t)$ & Noisy measurement of ${\delta}y$ (-)\\
$\hat{{\delta}y}(t)$ & Output deviation estimate (-)\\
${\delta}u(t)$ & Control input deviation from equilibrium (-)\\
${\Phi}_{yu}(\omega)$ & Output-input cross-spectral density (dB/Hz)\\
${\Phi}_{uu}(\omega)$ & Input power spectral density (dB/Hz)\\
${\psi}(f)$	  & Quasi-steady jet response (-)\\
${\psi}_D(f)$ & ${\psi}(f)$ at design flow rate (-)\\
$\gamma$ 	  & Chirp sweep rate (Hz/s)\\
$\xi$ 		  & Initial chirp frequency (Hz)\\
$\rho$ 		  & Density of air (kg/$\text{m}^3$)\\
${\sigma}^2_y$& Sensor noise variance for ${\delta}y'(t)$ (-)\\
$\mathbf{A}$  & State matrix (-)\\
$\mathbf{A}_\text{aug}$ & Augmented state matrix (-)\\
$b$		 & Inlet nozzle width (mm)\\
$d$		 & Inlet nozzle height (mm)\\
$\mathbf{B}$ & Input vector (-)\\
$\mathbf{B}_\text{aug}$ & Augmented input vector (-)\\
$\mathbf{C}$ & Output vector (-)\\
$\mathbf{C}_\text{aug}$ & Augmented output matrix (-)\\
$C(z)$   & Controller transfer function (-)\\
$\mathbf{e}(t)$	& State estimation error (-)\\
$\mathbf{e}_\text{aug}(t)$	& Augmented state estimation error (-)\\
$F(x)$ 	 & System nonlinearity (-)\\
$\mathbf{F}$ & Disturbance input vector (-)\\

\end{tabular}}
\end{minipage}
\begin{minipage}[t]{0.49\textwidth}
{\renewcommand{\arraystretch}{1.0}
\noindent\begin{tabular}{lp{6cm}}
\multicolumn{2}{l}{$\left.\right.$}\\

$\mathbf{F}_\text{aug}$ & Augmented disturbance input vector (-)\\
$f_0$	 & Dynamic response initial frequency (Hz)\\
$f_1(t)$ & Dynamic response step signal (Hz)\\
$f_m(t)$ & Modulating tone frequency (Hz)\\
$f_c(t)$ & Carrier tone frequency (Hz)\\
$f_s$	 & Sampling frequency (Hz)\\
$G(\omega)$       & Jet dynamic system transfer function (-)\\
$\hat{G}(\omega)$ & Empirical transfer function estimate (-)\\
$g_\text{amp}(t)$ & Audio amplifier signal (V)\\
$g_c(t)$ & Carrier tone signal (-)\\
$g_m(t)$ & Modulation signal / feed-back term (-)\\
$g_m'(t)$ & Control input (-)\\
$H(x)$	 & Limiting function (saturation limits) (-)\\
$H_\text{plant}(z)$ & Transfer function fitted to plant (-)\\
$K_\text{AW}$		& Anti wind-up gain (-)\\
$K_\text{DC}$ & System DC gain (-)\\
$K_\text{LQR}$ & LQR control gain vector (-)\\
$\dot{m}$	& Inlet mass flow rate (slpm)\\
$n$		 & Number of time series (-)\\
$N$	     & Number of segments (-)\\
${\Delta}P$ & Pressure difference across jet (Pa)\\
$r(t)$	 & Controller reference (-)\\
$R$ 	 & Jet radius of attachment (mm)\\
$T_s$    & Sampling time (s)\\
$S(z)$   & Sensitivity transfer function (-)\\
$u(t)$	 & Control input (-)\\
$u_0$ & Control input equilibrium (-)\\
$u_\text{FF}(t)$ & Feed-forward term in control input (-)\\

\end{tabular}}
\end{minipage}
\\
\begin{minipage}[t]{0.49\textwidth}
{\renewcommand{\arraystretch}{1.0}
\noindent\begin{tabular}{lp{7cm}}
	
\multicolumn{2}{l}{$\left.\right.$}\\

$\mathbf{v}(t)$ & Sensor noise (-)\\
$\mathbf{V}$		 & Sensor noise covariance matrix (-)\\
$W$		 & Process noise variance (-)\\
$w(t)$	 & Process noise signal (-)\\
$\mathbf{x}(t)$ & State vector (-)\\
$\mathbf{x}_\text{aug}(t)$ & Augmented state vector (-)\\
$\mathbf{x_0}$ & State vector equilibrium (-)\\
$X_\text{unforced}$ & Unforced DC output (-)\\
$y(t)$  & $i^\text{th}$ system output (-)\\
$\bar{y}(t)$ & Ensemble averaged output time series (-)\\
$z(t)$	 & Integrator state (-)\\
$z'(t)$	 & Noisy measurement of z(t) (-)\\
$\hat{z}$ & Integrator state estimate (-)\\
\\
\multicolumn{2}{l}{\textbf{Abbreviations}}\\
AM   & Amplitude modulation\\
BM   & Burst modulation\\

\end{tabular}}
\end{minipage}
\begin{minipage}[t]{0.49\textwidth}
{\renewcommand{\arraystretch}{1.0}
\noindent\begin{tabular}{lp{6cm}}
\multicolumn{2}{l}{$\left.\right.$}\\

ETFE & Empirical transfer function estimate\\
FFT  & Fast Fourier Transform\\
FPGA & Field programmable gate array\\
GM   & Gain margin\\
IMC  & Internal model control\\
LQR  & Linear quadratic regulator\\
LQG  & Linear quadratic gaussian\\
LTR  & Loop transfer recovery\\
LUT  & Look-up table\\
MC 1 & Measurement connection 1\\
MC 2 & Measurement connection 2\\
NMP  & Non-minimum phase\\
PM   & Phase margin\\
PT A & Pressure transducer A\\
PT B & Pressure transducer B\\
RMS  & Root mean square\\
ZNMF & Zero-net mass flux\\

\end{tabular}}
\end{minipage}

\doublespacing

\section{Introduction}
\label{sec:introduction}
This paper demonstrates closed-loop control of a novel piezo-fluidic amplifier that may be used in high-speed flow control applications \cite{stephens2009control, behr2007control, schabowski2014reduction}.
Although we present here the example problem of a fluidic diverter (see Fig. \ref{fig:device_diagram}) being used as an amplifier for a piezo actuator, the technique described in the paper can be used for control of any device based on fluidic jets.
Our work is motivated  by the challenge of controlling fluid flows in aerospace applications \cite{hak2000flow, jahanmiri2010active} that offers potential for fuel burn reduction through airframe \cite{corke_wing} as well as propulsion \cite{behr2007control,BacicVKI} performance improvements. A major challenge to the practical use of active flow control concepts \cite{cattafesta2011actuators} 
is the reliability, the authority, and relatively high bandwidth ($>100 \text{ Hz}$) required of the actuators for control of high-Reynolds number flows often encountered in aerospace applications. While actuation on $O(100 \text{Hz})$ is achievable with mechanical components like solenoid valves in benign low temperature environments, the same is not true for high-temperature applications such as jet engines \cite{BacicVKI}. While zero-net mass flux (ZNMF) actuators like piezoelectric buzzers may be practicable from a reliability perspective, they have limited authority in comparison with mechanical devices \cite{cattafesta2011actuators}. One way of addressing the bandwidth and authority requirements is through the use of passive fluidic oscillators \cite{osterman2008}. These have been utilised for a range of flow control applications from improving the effectiveness of a vertical tail \cite{Seele13,WhelanAIAA2018} to noise control \cite{raghu2004}. Another means to achieve both high bandwidth and high authority is by amplifying the effect of plasma or piezo actuators by switching fluidic diverters \cite{chen2016exp, mair2016experimental, mair2017switching, mair2018on}. However, unlike conventional flow control valves, neither of these approaches allows for controlled modulation of the output flow according to a desired reference trajectory. Piezo-fluidic amplifiers \cite{mair2016experimental,mair2017switching, mair2018on} rely on the fundamental physics of jet dynamics when subjected to sound injection in a transverse direction. The purpose of this paper is to demonstrate the effective closed-loop control of such dynamics thereby enabling continuous output flow modulation according to a desired output trajectory.
\\
\\
The physical understanding, modelling, and control of the nonlinear dynamics of fluids subject to external excitation is an important research topic.
Much of the published work in the field consists of numerical studies \cite{balogh2005optimal, balogh2001stability, aamo2013flow}, and more practical work typically comprises either open-loop control demonstrations \cite{jacobson1998active, bechert1975amplification, wiltse1993manipulation} or extremum-seeking schemes \cite{kegerise2004real, tian2006adaptive, becker2007adaptive, beaudoin2006drag}.
There are comparatively few examples of feedback controllers based on dynamic models \cite{henning2005drag, pinier2007proportional, rapoport2003closed, hecklau2011active}.
In \cite{wiltse1993manipulation}, a square jet was excited with piezoelectric actuators, which were driven with an amplitude modulated signal about their resonant carrier frequency. This study demonstrated that the jet demodulated the excitation signal and responded at the modulation frequency. This result has been used to control jets in several studies since \cite{tian2006adaptive,rapoport2003closed} and is also used in the research presented here. For example, in \cite{tian2006adaptive}, an adaptive closed-loop control scheme is used to explore the optimal AM or BM (burst modulated) forcing frequency of a synthetic jet actuator, which injected flow into the boundary layer of a NACA airfoil to promote reattachment of the separated flow. A strain gauge was used to determine the lift and drag forces on the airfoil, which were used to assess the degree of separation in the cost function for the adaptive control algorithm. This approach yielded a doubling of the lift-to-drag ratio.
\\
\\
In \cite{henning2005drag}, the drag over a bluff body was reduced by excitation with synthetic jet actuators on the trailing edge to suppress the natural vortex shedding frequency. A first order model with a static nonlinear map and delay was used for synthesizing a robust controller. The controller was able to track a reference pressure coefficient whilst the flow Reynolds number varied significantly, highlighting the benefits of using feedback. A second control scheme based on an extremum-seeking method made use of an extended Kalman filter to estimate amplitude, frequency, and phase of the fluctuations in the flow. This controller was found to achieve the same recovery of pressure and reduction of drag for half the actuation energy as the first scheme.
\\
\\
In \cite{rapoport2003closed}, a jet issuing from a nozzle terminated with a wide-angle diffuser was encouraged to attach to the diffuser by thrust vectoring using a synthetic jet actuator. The main jet responded to the modulation signal of an AM-driven synthetic jet. System identification was used to fit a second order dynamic model and a controller was designed using an internal model control (IMC) scheme. The controlled jet responded up to 30 - 50 Hz, achieving an order of magnitude higher bandwidth than conventional thrust vectoring mechanisms.
\\
\\
In this paper, we employ the same piezo actuation method as in \cite{mair2016experimental} where a highly reliable piezo buzzer is used to deflect the jet  inside a Coand{\u{a}} diverter causing the device to switch its state. However, the response of the device is limited in Mair et al. \cite{mair2016experimental, mair2017switching, mair2018on} by the application of open-loop control only.  The use of open-loop control has two main disadvantages: $(i)$ no disturbance rejection, and $(ii)$ significant variation of the switching time. The former makes the system susceptible to upstream and downstream pressure changes whereas the latter precludes the use of the device in applications that require strong synchronisation. In this paper we propose the use of closed-loop control to tackle both of these deficiencies by utilising pressure measurements from the total pressure tappings in the output channels as feedback signals (see Fig. \ref{fig:device_diagram}) and by relying on the demodulating properties of jet dynamics as first demonstrated by Wiltse et al. \cite{wiltse1993manipulation}. Section \ref{sec:piezo-fluidic concept} reviews the piezo-fluidic concept. Section \ref{sec:system identification} describes the experimental set-up and carries out system identification of the jet dynamics inside the device. Section \ref{sec:controller design and simulation} describes the design of the output LQG controller. Finally, Section \ref{sec:implementation and results} demonstrates experimentally the effectiveness of the closed-loop scheme in controlling the jet.

\section{Piezo-Fluidic Concept}
\label{sec:piezo-fluidic concept}
\begin{figure*}[t]
\centering
\captionsetup{justification=centering}
\includegraphics[width=\textwidth]{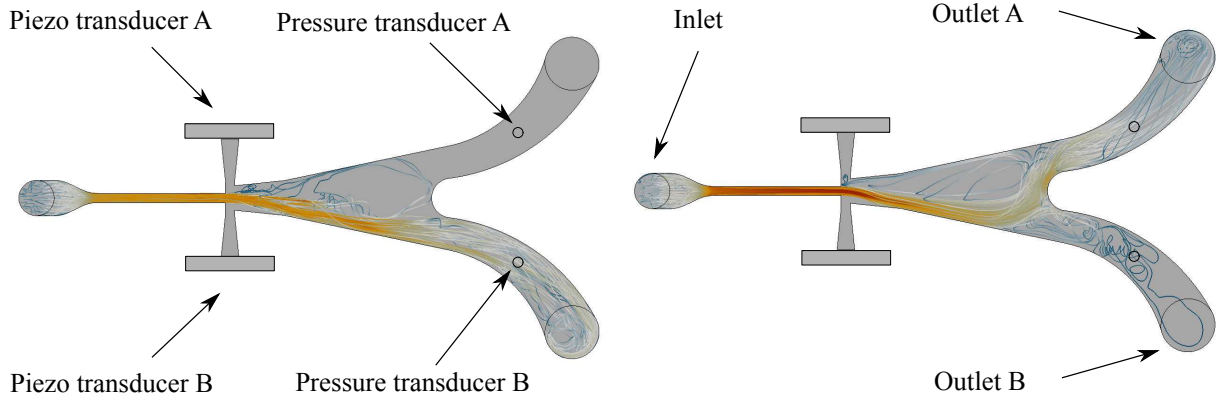}
\caption{\label{fig:device_diagram}Fluidic amplifier used in this paper.}
\end{figure*}

\begin{figure}[t]
\centering
\captionsetup{justification=centering}
\includegraphics[width=0.65\textwidth]{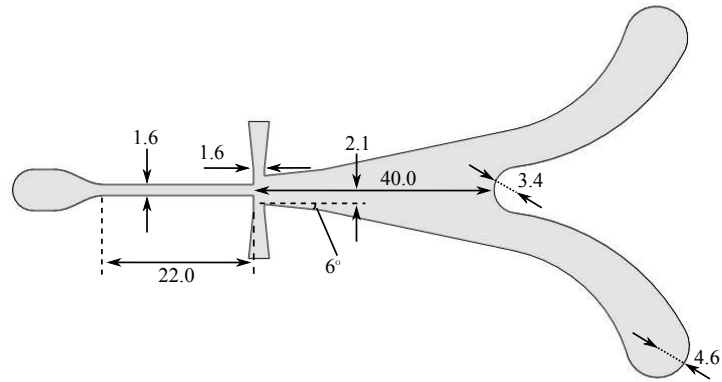}
\caption{\label{fig:device dimensions}Device dimensions. Unless otherwise indicated, the units are mm. The depth of the fluid path is $d = 4.8$ mm.}
\end{figure}

The device used is shown in Fig. \ref{fig:device_diagram}, with detailed dimensions given in Fig. \ref{fig:device dimensions}.
The inlet channel has a rectangular cross-section with width $b = 1.6$ mm by height $d = 4.8$ mm.
Fluidic amplifiers operate using the Coand{\u{a}} effect, which is the tendency of a jet to attach itself to a nearby surface \cite{tritton1977physical}. When supplied with a pressurised fluid, a jet issues into the interaction region of the device from the nozzle orifice and entrains the surrounding stationary fluid, which lowers the surrounding pressure. Any asymmetry causes the jet to bend to the side with slightly lower pressure, which acts to increase the pressure difference across the jet by confining the side towards which the jet bends and relieving the opposite side. The jet eventually strikes the wall to enclose a low-pressure separation bubble, which sucks in fluid from the jet that has insufficient total pressure to continue downstream \cite{chapman1957investigation}, counteracting the pressure-reducing effect of the entrainment and stabilising the bubble pressure. A steady state is reached once there is equilibrium between the entrainment and recirculation flows - this is known as the Coand\u{a} effect \cite{kirshner2012design}, which gives all wall-attachment devices their stability and supports a pressure difference across the jet. This results in a bistable device, with flow attaching to one of the two walls and exiting through the corresponding outlet channel.
Classically, such a diverter can be switched by injection or extraction of the transverse mass flow, causing the main jet to move past the splitter (which is set back and blunt in this device) and attach to the opposite wall. Recently, however, Mair et al \cite{mair2016experimental, mair2017switching} demonstrated that zero-net-mass-flux piezoelectric transducers can be used to switch a similar device using open-loop control at a characteristic frequency. The mechanism causing the jet deflection and subsequent switching depends on which side of the device the acoustic signal from the piezoelectric transducer (the excitation) is applied from relative to which channel the jet is attached to. It is applied either from the same side or the opposite side to the channel to which the jet is attached. In \cite{mair2016experimental} it is shown that, when exciting from the unattached side, the deflection is caused by exciting the shear layer roll-up mode of the jet, or one of its subharmonics. This is the most unstable natural mode, and has been shown in several studies to be at a nondimensional frequency of $St_{\theta} = 0.012$, e.g. \cite{zaman1980vortex, shih1992experimental}. However,  Mair et al. \cite{mair2016experimental} note that switching is possible across a broad range of frequencies when the flow rate is sufficiently low.
\\
\\
Exciting the jet shear layer results in a steady deflection of the jet so that a greater portion of it exits via the unattached side outlet. To understand this, consider the steady jet deflection equation, which is easily derived by considering the radial acceleration of a curved jet \cite{ries1972dynamic}
\begin{equation}
\label{eq: steady-state jet deflection equation}
\left(\frac{\dot{m}^2}{{\rho}bd^2}\right)\frac{1}{R} \, = \, {\Delta}P,
\end{equation}
where $\dot{m}$ is the jet mass flow, $b$ and $d$ are the width and height of the nozzle cross-section from which the jet emerges respectively, $R$ is the radius of attachment of the jet, and ${\Delta}P$ is the pressure difference across the jet. The radius of attachment, $R$, results from assuming that the jet centreline follows the arc of a circle between the nozzle orifice and the attachment point (where it strikes the wall) \cite{bourque1960reattachment}. Therefore, the smaller the value of $R$, the tighter the attachment and the greater pressure difference across the jet, ${\Delta}P$, is required to supply the centripetal acceleration to maintain the arc shape. This equation describes the strength of the Coanda effect - smaller values of ${\Delta}P$ and larger values of $R$ for a given mass flow rate, $\dot{m}$, indicate that the Coanda effect is weaker and the jet less firmly attached to the wall. Exciting the shear layer promotes vortex production, which increases the entrainment flow on both sides of the jet, but more so on the unattached side \cite{mair2016experimental}. This results in a biased reduction in pressure on either side of the jet, such that ${\Delta}P$ decreases in magnitude. Therefore the radius of attachment, $R$, increases, the Coand\u{a} effect is weakened, and the jet is deflected away from the wall. In the limiting case that $R \to \infty$, i.e. a straight jet, the flow would be divided equally between the two outlet channels by the splitter. Hence, increasing $R$ by acoustically exciting the unattached side shear layer results in a portion of the jet exiting the fluidic device via the unattached side outlet.
\\
\\
In this paper, we propose the idea of using a carrier tone to which amplitude modulation is applied as a control signal. Therefore, our control signal $u(t)$ will amplitude modulate the carrier tone, $u_\text{piezo}(t) = u(t)sin(\omega_ct)$. The flow rate used in this paper was 40 lpm, corresponding to a mean inlet channel velocity of 90 ms$^{{\minus}1}$, an inlet-to-outlet pressure ratio of 1.1, and a Reynolds number based on the hydraulic diameter of 2$.$2 $\times$ 10$^\text{4}$. At this flow rate, it is not possible to make the jet switch, although large deflections of the jet are possible, and a large portion of the flow can be directed out of the unattached side channel.
Hence, we aim to provide a second mode of operation for the device. This mode operates at higher jet speeds relative to conventional operation so that we obtain a faster response, but due to limited piezo amplitudes does not lead to a full switch. This results in lower effective gain (from piezo amplitude to total pressure in the unattached-side outlet channel) as only part of the jet is directed out of the unattached side outlet. However, since the jet is never fully detached, the slow dynamics of detachment and reattachment are avoided, details of which can be found in Epstein \cite{epstein1971theoretical}. This, in combination with higher jet speeds ensures, a faster device response, leading to higher effective bandwidth. The output used is the total pressure in the unattached side channel, as measured by a total pressure tapping in the centre of the channel. While it would be more accurate to use, for example, a hot-wire anemometer at the unattached side outlet orifice, this measurement strategy would not be sufficiently robust to be used in a real application.
\section{System Identification}
\label{sec:system identification}
\subsection{Experimental Set-up}
\begin{figure}
\centering
\captionsetup{justification=centering}
\includegraphics[width=0.5\textwidth]{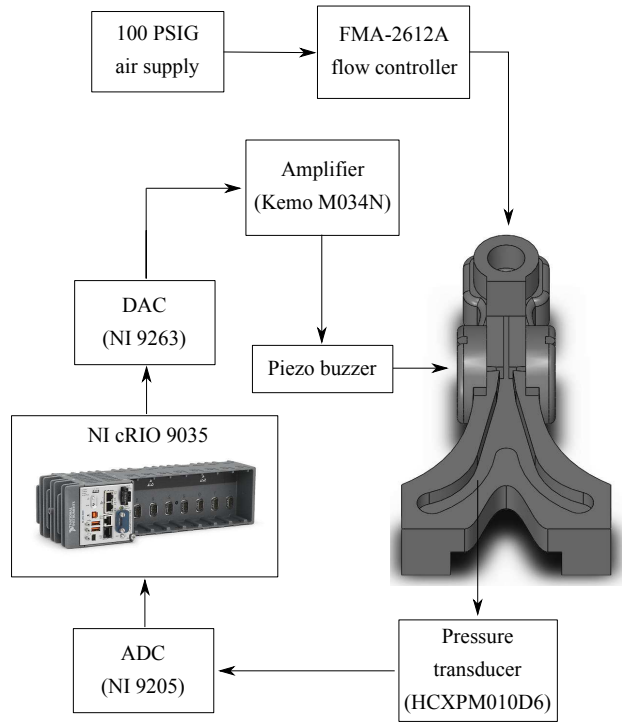}
\caption{\label{fig:experimental setup}Experimental set-up}
\end{figure}

\begin{figure}
\centering
\captionsetup{justification=centering}
\includegraphics[width=0.7\textwidth]{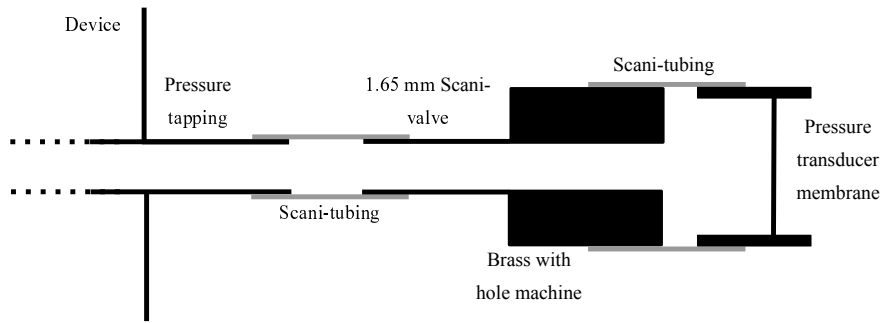}
\caption{\label{fig:measurement connection}Measurement connection diagram}
\end{figure}

The experimental set-up is shown in Fig. \ref{fig:experimental setup}.
The FPGA (field programmable gate array) used is the National Instruments (NI) cRIO-9035, with the NI 9205 Analogue Input (AI) card and NI 9263 Analogue Output (AO) card. The piezo used is the Kingstate 108 dB Panel Mount Continuous External Piezo Buzzer, the audio amplifier is the Kemo 40 W M034N, and the pressure transducer is the First Sensor 10 mbar HCXPM005D6V (henceforth referred to as PT A), which has a response time of 100 ${\mu}$s, and the Kulite XCQ-series pressure transducer (PT B). PT A was used in all cases except when stated otherwise. A measurement connection (MC) was required to connect PT A to the total pressure tapping on the device. This consisted of a short length of 1.65 mm Scanivalve tubing to connect the tapping to an adapter, which comprised a short length of 1.6 mm Scanivalve soldered to a thicker brass cylinder with the same internal diameter. A second piece of Scanivalve tubing of appropriate (larger) diameter connected the adapter to the pressure transducer. A diagram of the measurement connection is shown in Fig \ref{fig:measurement connection}.
There were two versions of this measurement connection: MC 1 and MC 2. The latter version simply had the shortest possible lengths of each section of the connection, thus reducing its filling time and increasing its bandwidth. The First Sensor pressure transducer has a response time of 100 $\mu$s. However, the limiting factor in the measurement frequency response is the filling associated with the measurement connection. Another pressure transducer, a Honeywell SDX series device, was also used but showed no improvement in the frequency response for the same reason. It is the measurement connection that causes the roll-off.
\\
\\
The dynamics that we have studied in the present work are those of the bulk jet rather than those associated with shear layer instabilities. This would usually inform the choice of sampling rate, however in our case it was necessary to generate input waveforms, and it was convenient to match the controller loop rate to the sampling rate. The frequency of the input signals required was $O(1 \text{kHz})$, and to synthesize these signals accurately we required a loop rate at least one order of magnitude greater. The loop and sampling rates were therefore set to $f_s = 50$ kHz in all cases. An analogue, first order, 25 kHz (the Nyquist frequency) anti-aliasing filter was applied to the pressure transducer measurements before sampling by the FPGA, and the flow controller used was the Omega FMA-2612A. The frequency resolution of signals obtained varied from 30 Hz for the ETFEs in Section \ref{sec:dynamic ID} to 0.02 Hz for the quasi-steady jet response obtained from simulations, also in Section \ref{sec:dynamic ID}. For stationary signals, a standard frequency-domain averaging method was used, where sampled time series were split into segments, their FFTs were calculated then the collection of FFTs were averaged \cite{Ljung}. The number of segments (N) used determined the frequency resolution ($f_s/N$), which varied depending on the signal processing task and was chosen to keep the signal-to-noise ratio above 10 dB.
\subsection{Linearity}
\label{sec:linearity}
\begin{figure}
\centering
\captionsetup{justification=centering}
\includegraphics[width=0.5\textwidth]{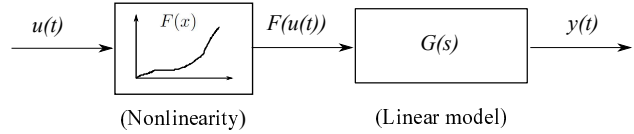}
\caption{\label{fig:hammerstein model} Approximation of plant: Hammerstein model}
\end{figure}

\begin{figure}[h!]
\centering
\captionsetup{justification=centering}
	\begin{subfigure}{0.32\textwidth}
	\includegraphics[width=\textwidth]{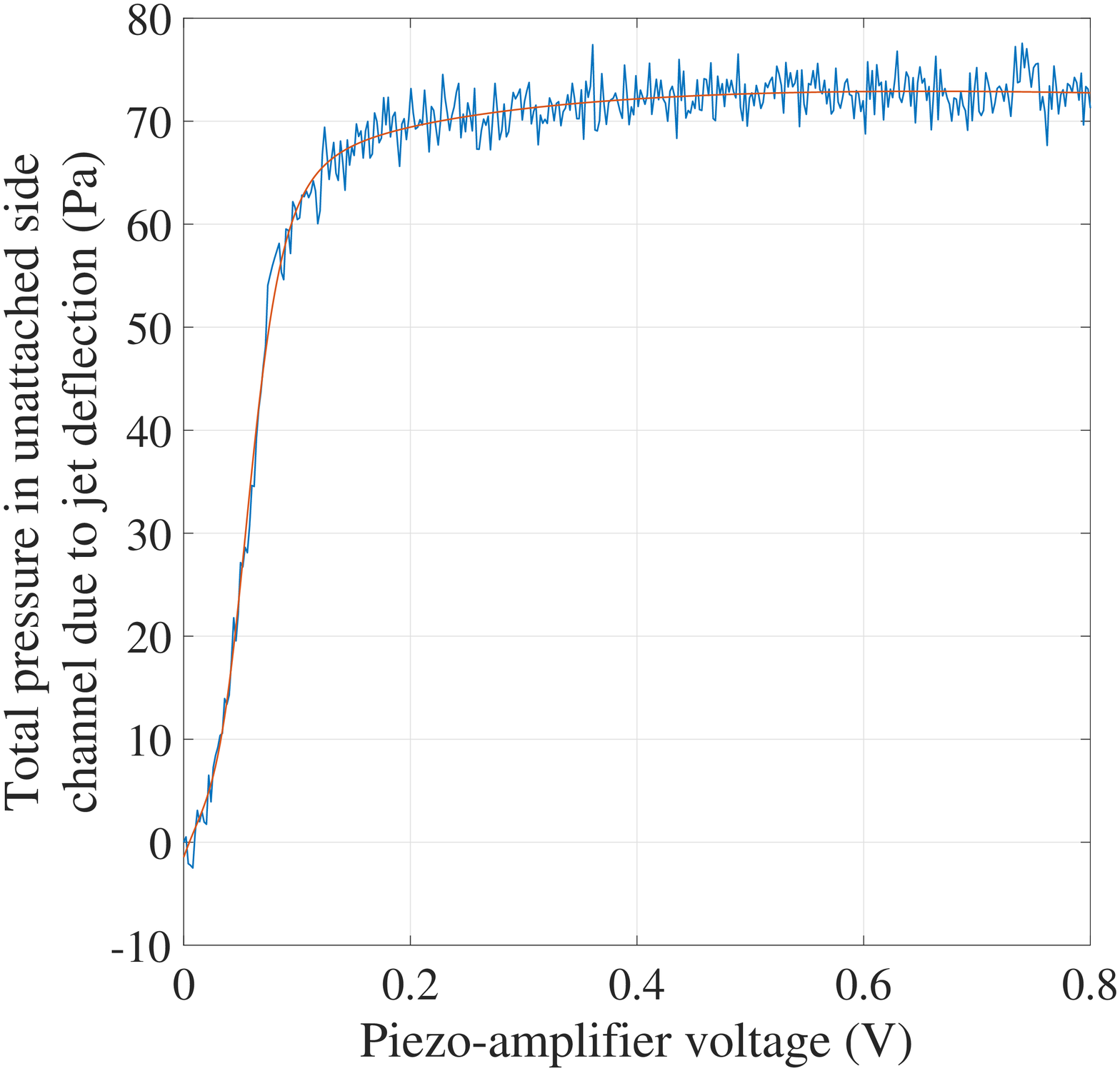}
	\caption{\label{fig:30lpm NL}30 lpm}
	\end{subfigure}
	\begin{subfigure}{0.32\textwidth}
	\includegraphics[width=\textwidth]{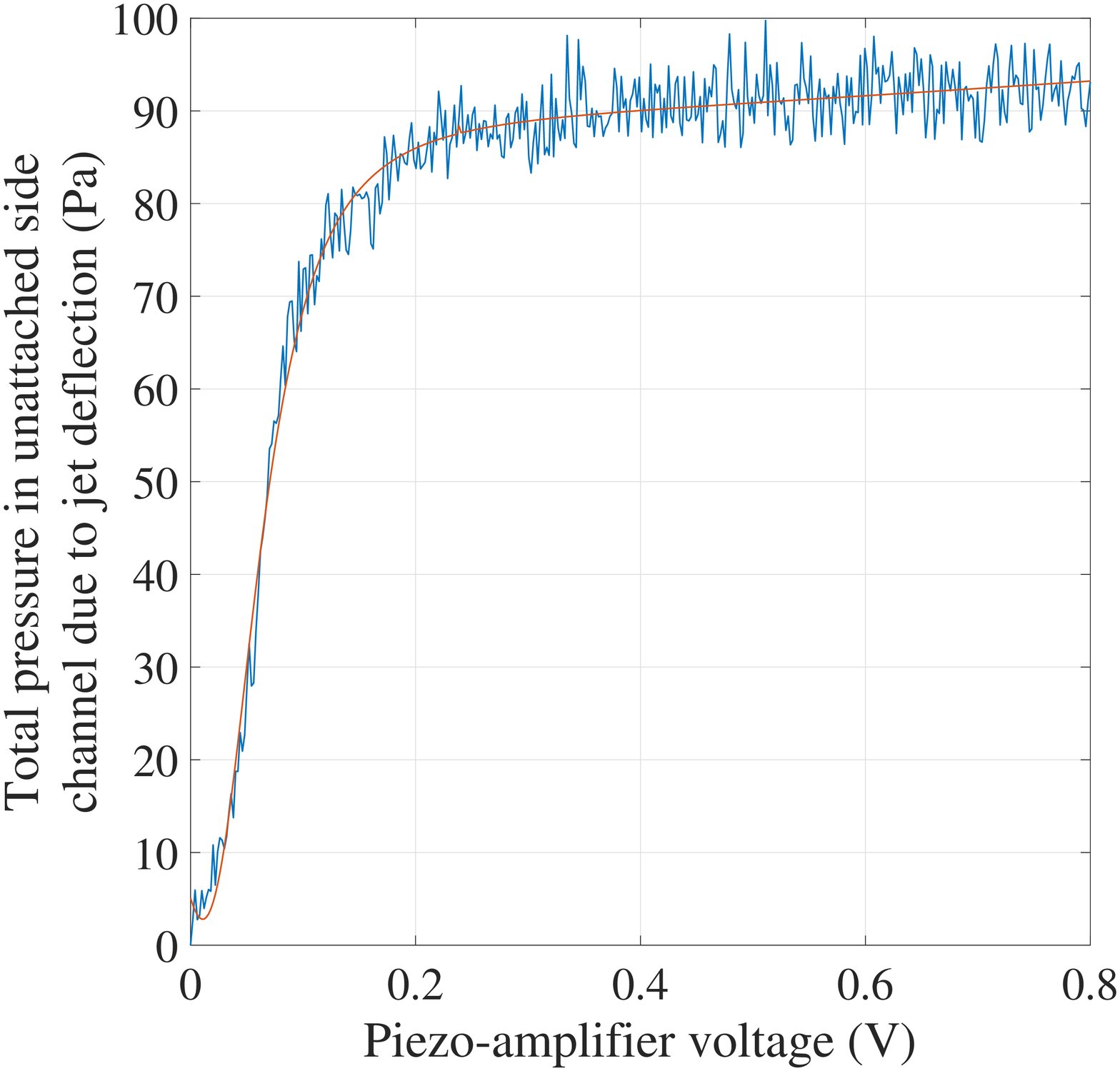}
	\caption{\label{fig:35lpm NL}35 lpm}
	\end{subfigure}
	\begin{subfigure}{0.32\textwidth}
	\includegraphics[width=\textwidth]{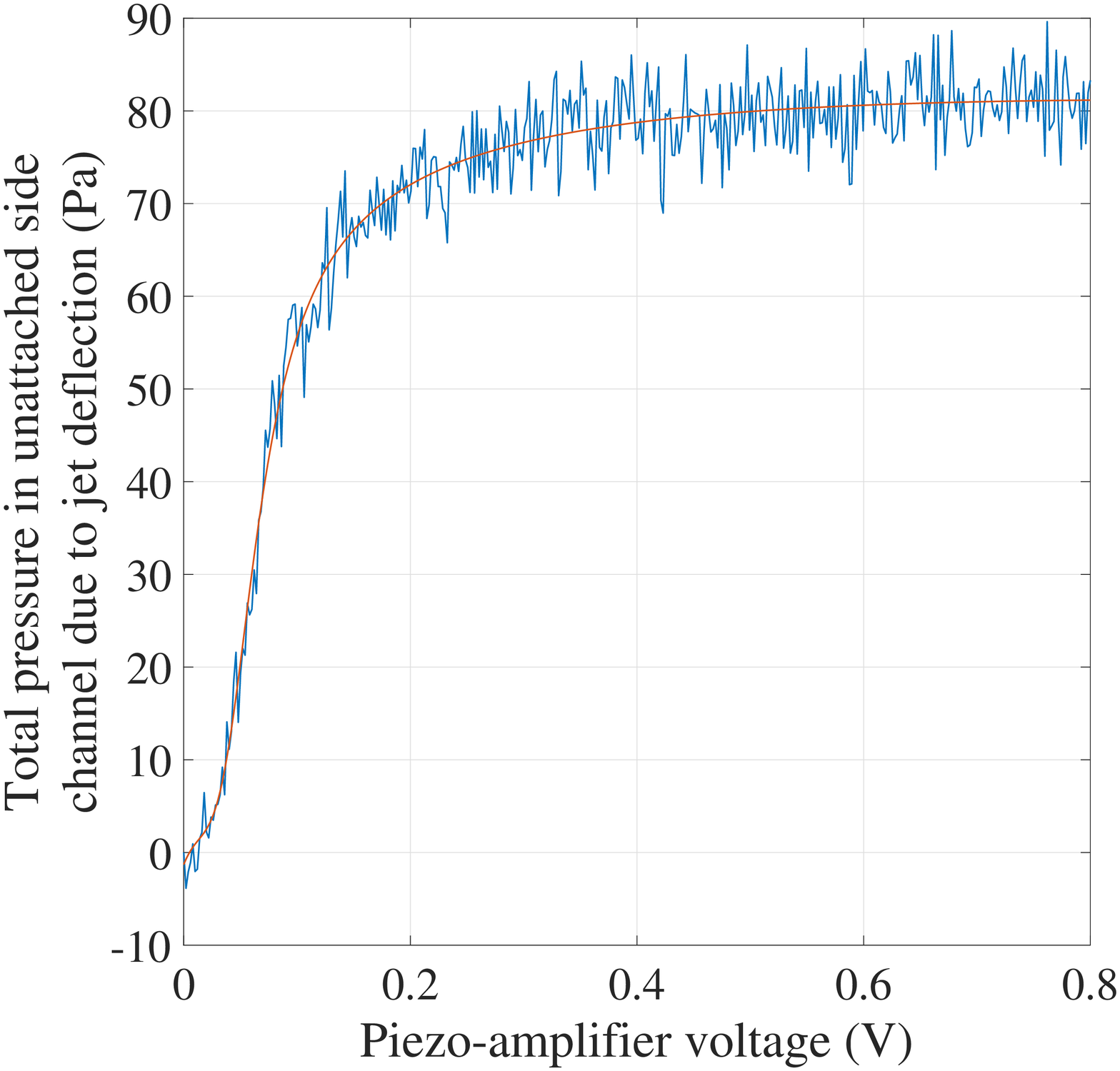}
	\caption{\label{fig:40lpm NL}40 lpm - design case}
	\end{subfigure}
	\begin{subfigure}{0.32\textwidth}
	\includegraphics[width=\textwidth]{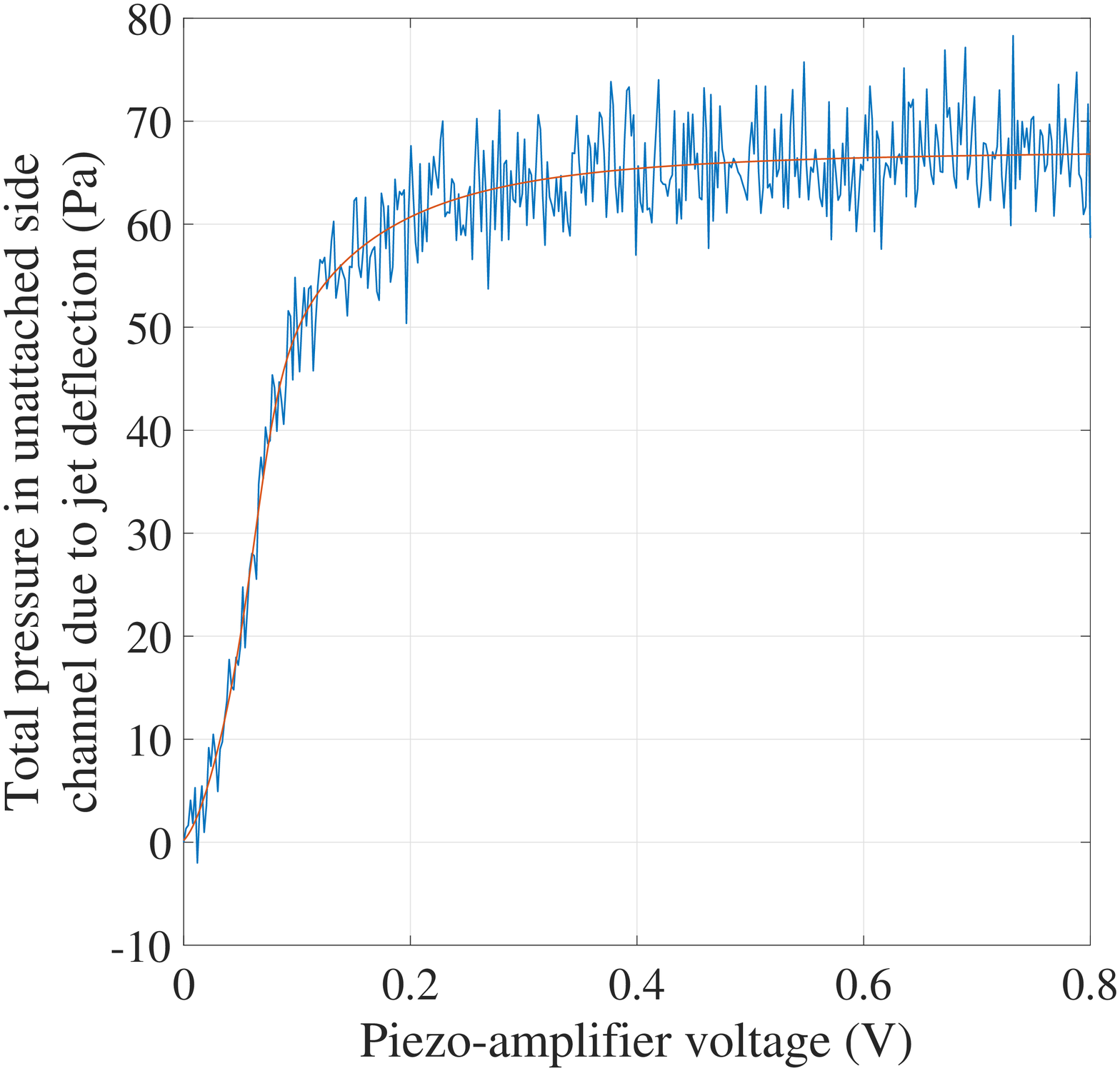}
	\caption{\label{fig:45lpm NL}45 lpm}
	\end{subfigure}
	\begin{subfigure}{0.32\textwidth}
	\includegraphics[width=\textwidth]{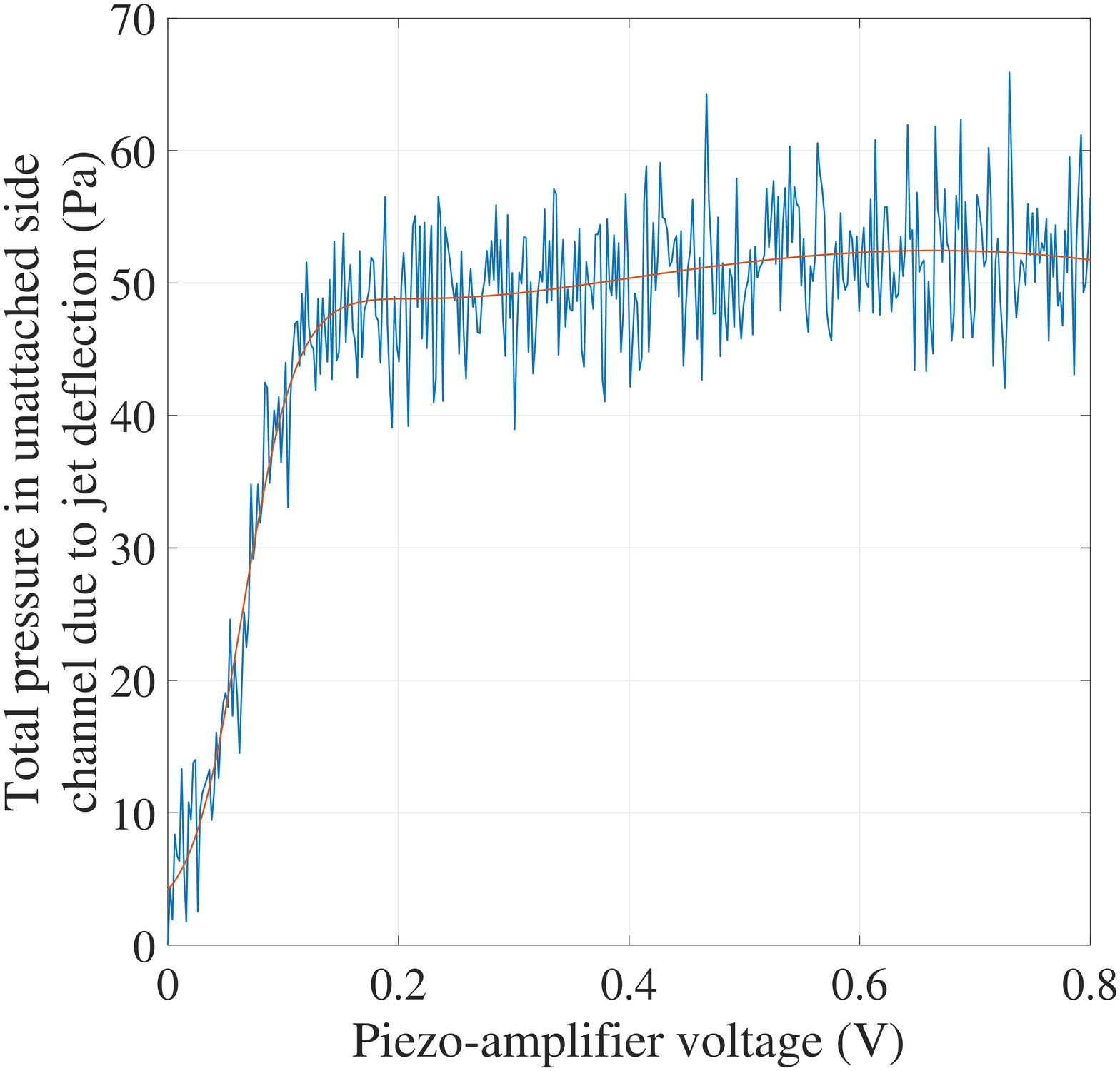}
	\caption{\label{fig:50lpm NL}50 lpm}
	\end{subfigure}
	\begin{subfigure}{0.32\textwidth}
	\includegraphics[width=\textwidth]{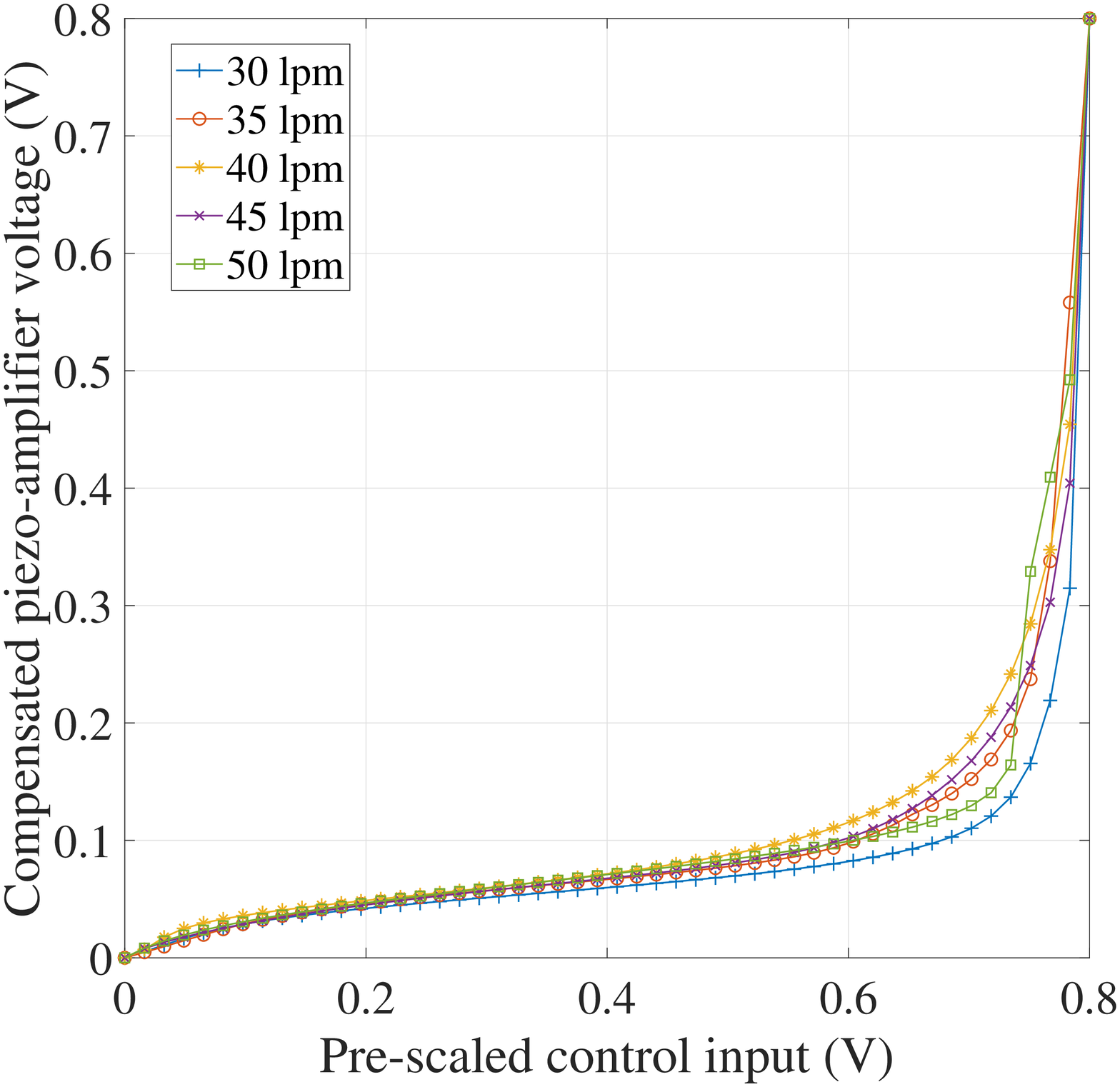}
	\caption{\label{fig:NLs inverted LUT data}Scaled, inverted functions}
	\end{subfigure}
	\caption{\label{fig:NL data}System nonlinearities, $F(x)$, at several flow rates: ensemble averaged data (N = 2, blue) and fitted rational functions (red), $f_c =$ 2.75 kHz; Fig. \ref{fig:NLs inverted LUT data} shows the scaled, inverted functions for each flow rate for look-up table implementation.}
\end{figure}

\begin{figure}
\centering
\captionsetup{justification=centering}
\includegraphics[width=0.6\textwidth]{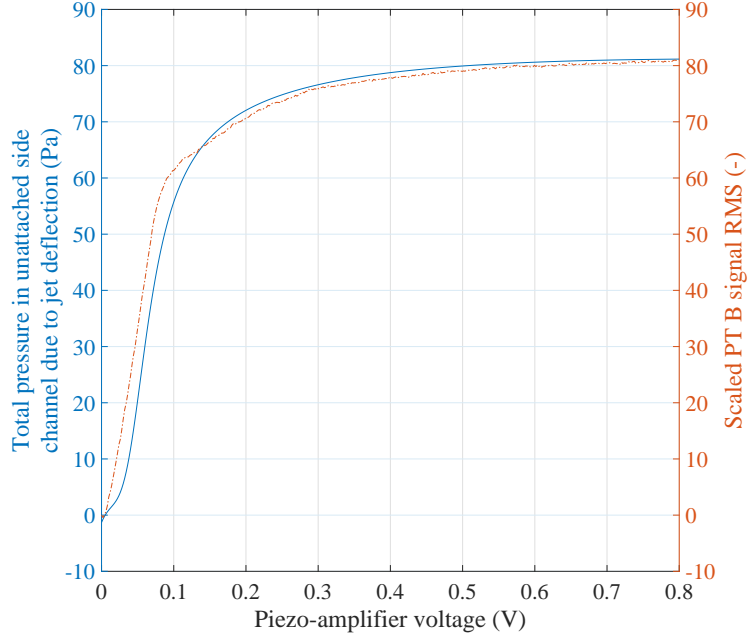}
\caption{\label{fig:IO NL and input NL}Input-output nonlinearity rational function at 40 lpm (solid blue) and scaled input nonlinearity as measured by PT B (Kulite) RMS (dash-dot red)}
\end{figure}

There are several sources of nonlinearity. The pressure tapping measures the pressure across a central section of the unattached side channel. This is in the shear layer of the deflected jet, which does not have a linear velocity profile, so that as the deflection varies the pressure measurement varies nonlinearly.
The piezo-amplifier system is another source of nonlinearity. Finally, the jet response to excitation is faster than the speed of the jet's movement back to its natural, unexcited steady state. The system model can be approximated with a Hammerstein model, shown in Fig. \ref{fig:hammerstein model}, where the system nonlinearities have been incorporated into the static nonlinearity, $F(x)$.
\\
\\
The function $F(x)$ was determined at several flow rates by driving the audio amplifier at 2.75 kHz with amplitude linearly increasing from 0 to 10 V$_\text{pp}$. The resulting characterisations are shown in Fig. \ref{fig:NL data} with corresponding fitted rational functions. Input amplitudes greater than ${\sim}$ 0.8 V are omitted because the resulting deflection was not strictly monotonically increasing.
The curves in Fig. \ref{fig:NLs inverted LUT data} are the scaled, inverted, fitted functions that were implemented in look-up tables for the dynamic system identification experiments in Section \ref{sec:dynamic ID} in order to preserve linearity. These curves demonstrate that the function $F(x)$ is relatively insensitive to flow rate, suggesting that any feedback controller will not be limited in tracking reference jet positions away from the design flow rate because of a variation in the system nonlinearity.
\\
\\
An experiment was conducted in order to determine the significance of the contributions to the nonlinearity curves in Fig. \ref{fig:NL data}, which represent the overall input-output nonlinearity at each flow rate. A 2.75 kHz tone was used to drive the piezo-amplifier system, linearly increasing in amplitude from 0 to 0.8 V$_\text{pp}$ over 50 seconds, and the resulting sound pressure level was measured by PT B (the Kulite) in one of the outlet channels. The time series was split into 400 sets of 6250 samples, and the RMS was taken of each set, resulting in a 400-sample curve. These data were scaled up to the amplitude of the 40 lpm input-output nonlinearity rational function so as to draw a comparison, and both of these curves are shown in Fig. \ref{fig:IO NL and input NL}.
The RMS full scale error (i.e. error relative to the maximum value) of input nonlinearity (red) relative to the overall nonlinearity (blue) is 4.2\%. This justifies our use of a Hammerstein model structure, which assumes that the system nonlinearity is entirely at the input to the system (Fig. \ref{fig:hammerstein model}).
\subsection{Frequency Sensitivity of the Jet to Perturbation}
\label{sec:static ID}
\begin{figure}
\centering
\captionsetup{justification=centering}
\includegraphics[width=0.6\textwidth]{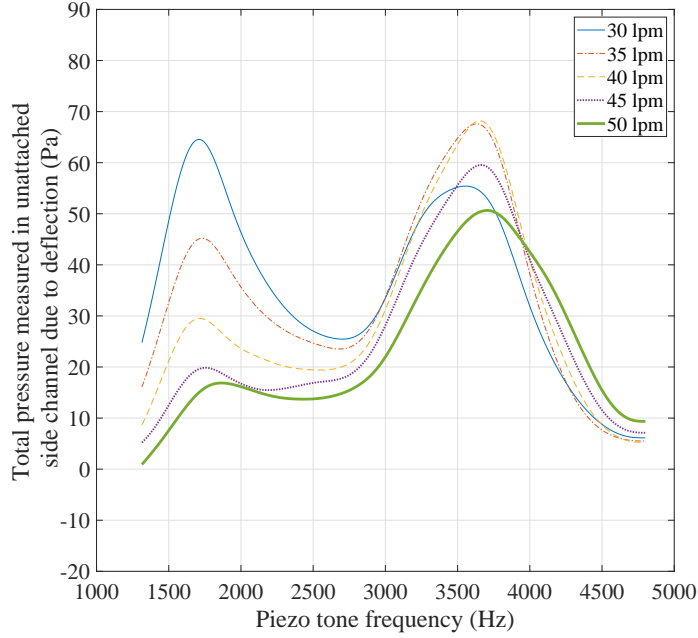}
\caption{\label{fig:static deflection curves}Static jet deflection vs perturbation tone frequency at several flow rates: 30 lpm (blue, solid), 35 lpm (red, dash-dot), 40 lpm (yellow, dash), 45 lpm (purple, dot), 50 lpm (green, bold solid)}
\end{figure}

When the piezo is driven by a single frequency tone in the range 1.3 to 4.8 kHz, the jet is deflected, resulting in a steady state increase in the unattached side total pressure. 
To identify frequencies at which the piezo-jet system is most responsive, the piezo on the unattached side was driven through the audio amplifier with a chirp input signal with an amplitude of 70 mV$_\text{pp}$ from 1.3 to 4.8 kHz over 100 s, i.e.
\begin{equation}
\label{eq:chirp signal for static sys ID}
u(t) \, = \, Asin\left[2\pi\left({\xi}t + \frac{{\gamma}}{2}t^2\right)\right],
\end{equation}
where $\gamma =$ 35 Hzs$^{-1}$, and $\xi =$ 1.3 kHz. The signal mean was sampled to determine the degree of deflection. Fig. \ref{fig:static deflection curves} shows the deflection curves.
It should be noted that this curve shows deflections away from the natural bias, which is itself a function of flow rate.
\subsection{Jet Dynamics}
\label{sec:dynamic ID}
\begin{figure}
\centering
\captionsetup{justification=centering}
\includegraphics[width=0.6\textwidth]{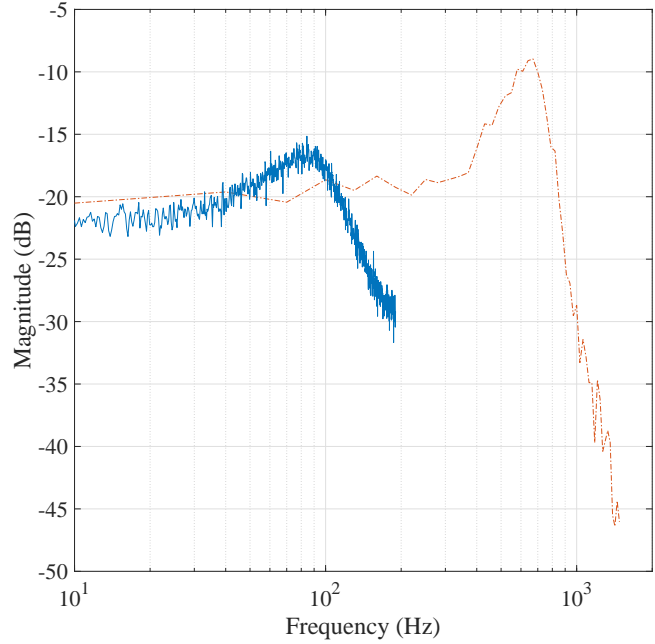}
\caption{\label{fig:dynamic system ID1} Open-loop Bode magnitude plot from ETFE at 40 lpm and $f_c \, = \, 2.75$ kHz. ETFE from MC 1 (blue, solid) and from MC 2 (red, dash-dot).}
\end{figure}

\begin{figure}[h!]
\centering
\captionsetup{justification=centering}
	\begin{subfigure}{0.32\textwidth}
	\includegraphics[width=\textwidth]{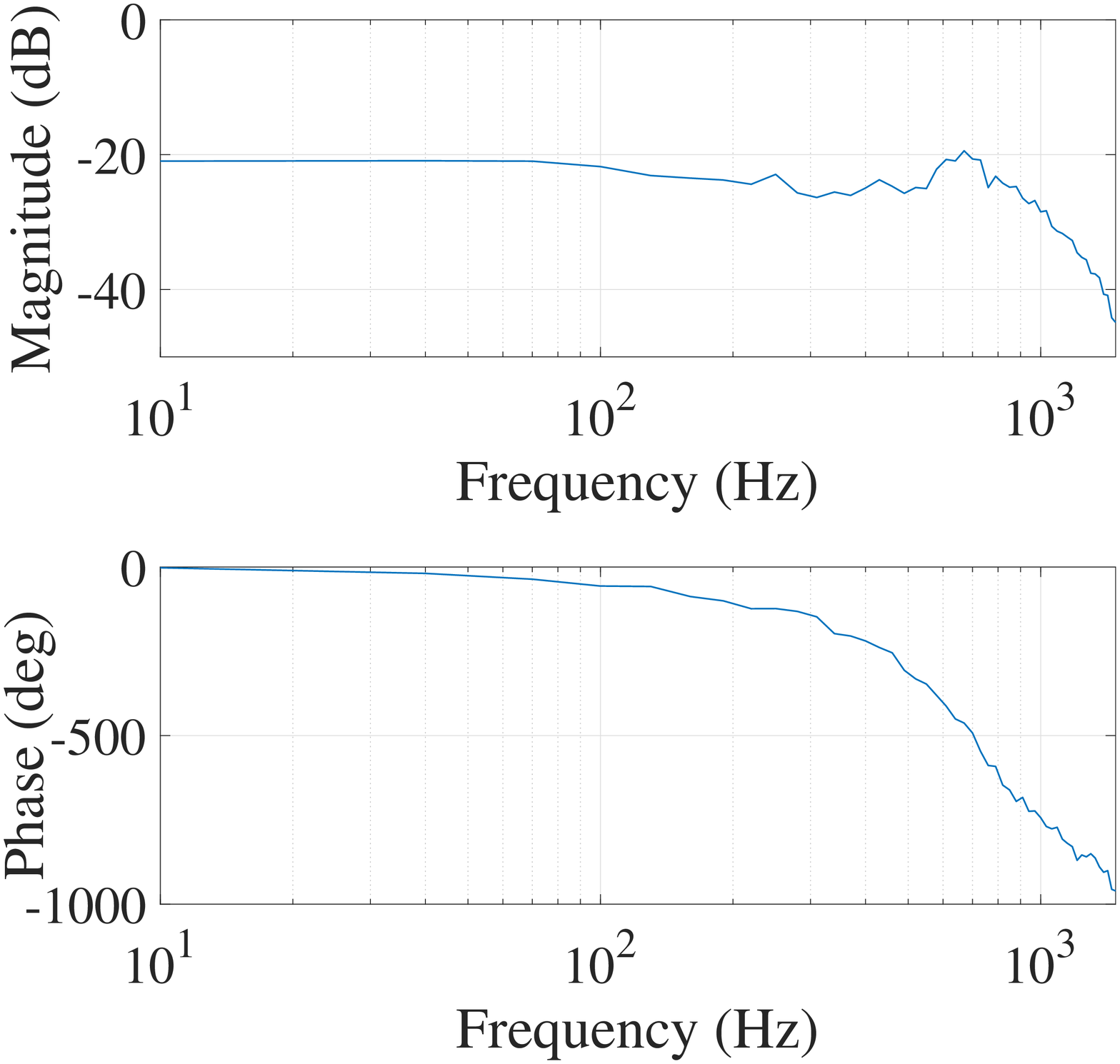}
	\caption{\label{fig:20lpm f resp}30 lpm}
	\end{subfigure}
	\begin{subfigure}{0.32\textwidth}
	\includegraphics[width=\textwidth]{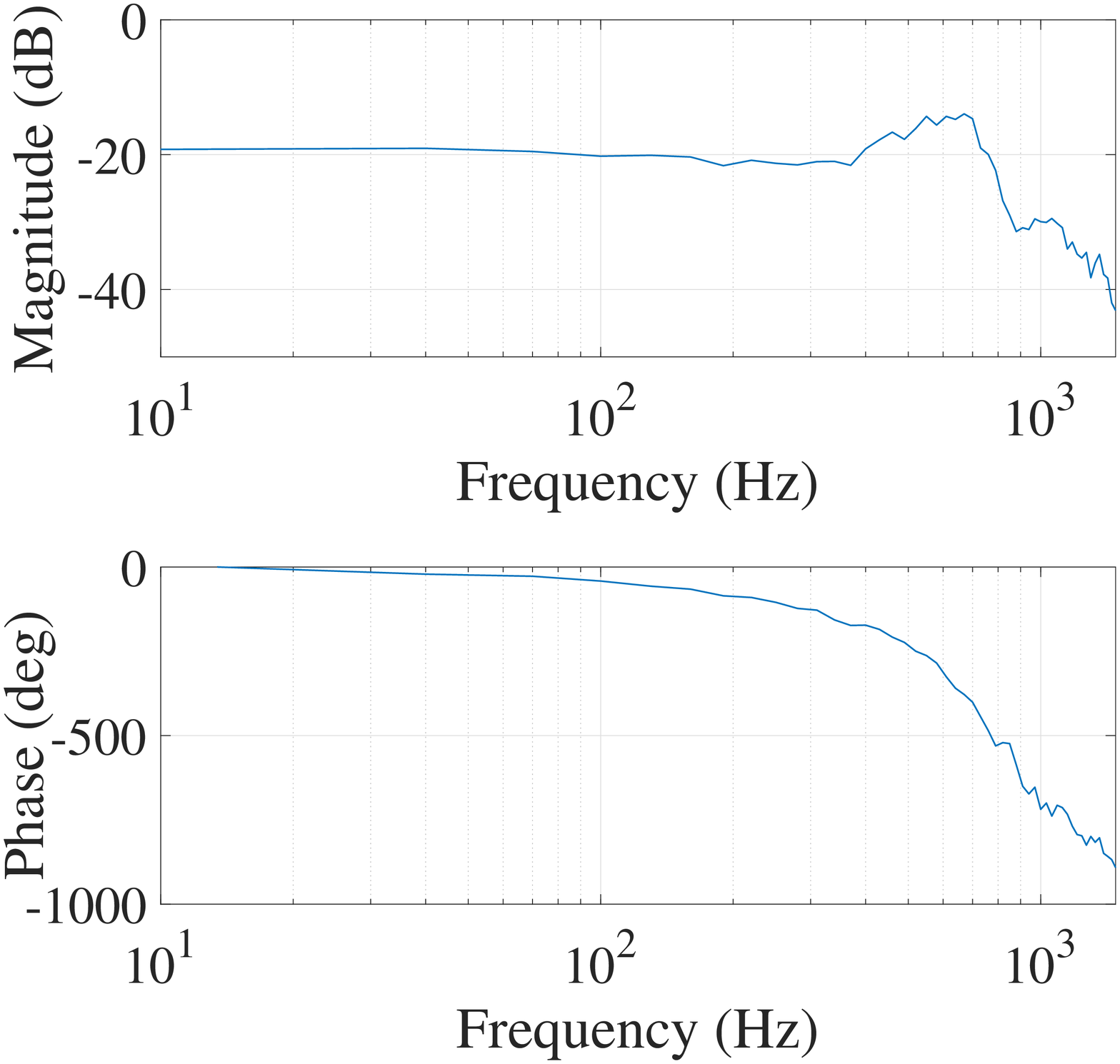}
	\caption{\label{fig:25lpm f resp}35 lpm}
	\end{subfigure}
	\begin{subfigure}{0.32\textwidth}
	\includegraphics[width=\textwidth]{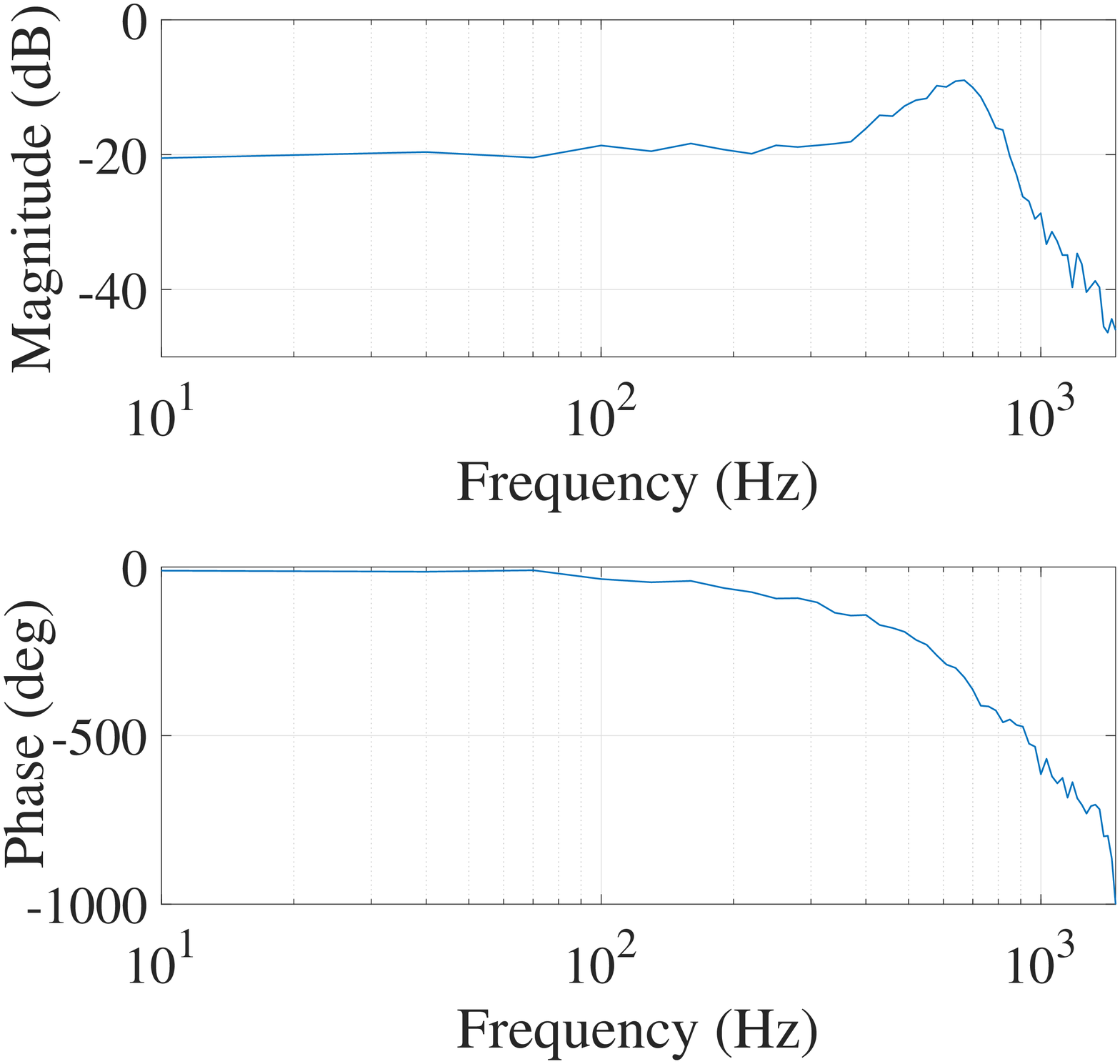}
	\caption{\label{fig:30lpm f resp}40 lpm - design case}
	\end{subfigure}
	\begin{subfigure}{0.32\textwidth}
	\includegraphics[width=\textwidth]{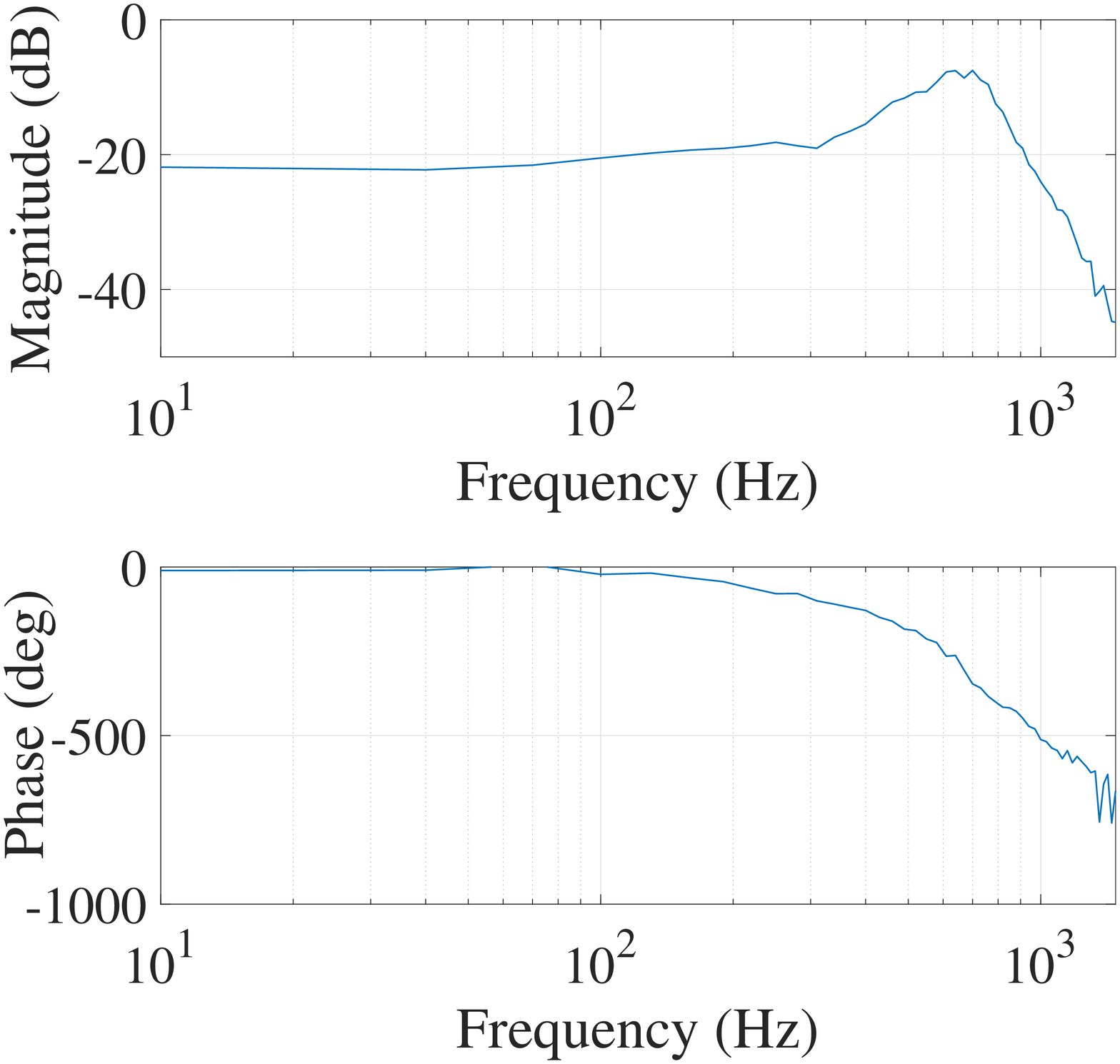}
	\caption{\label{fig:35lpm f resp}45 lpm}
	\end{subfigure}
	\begin{subfigure}{0.32\textwidth}
	\includegraphics[width=\textwidth]{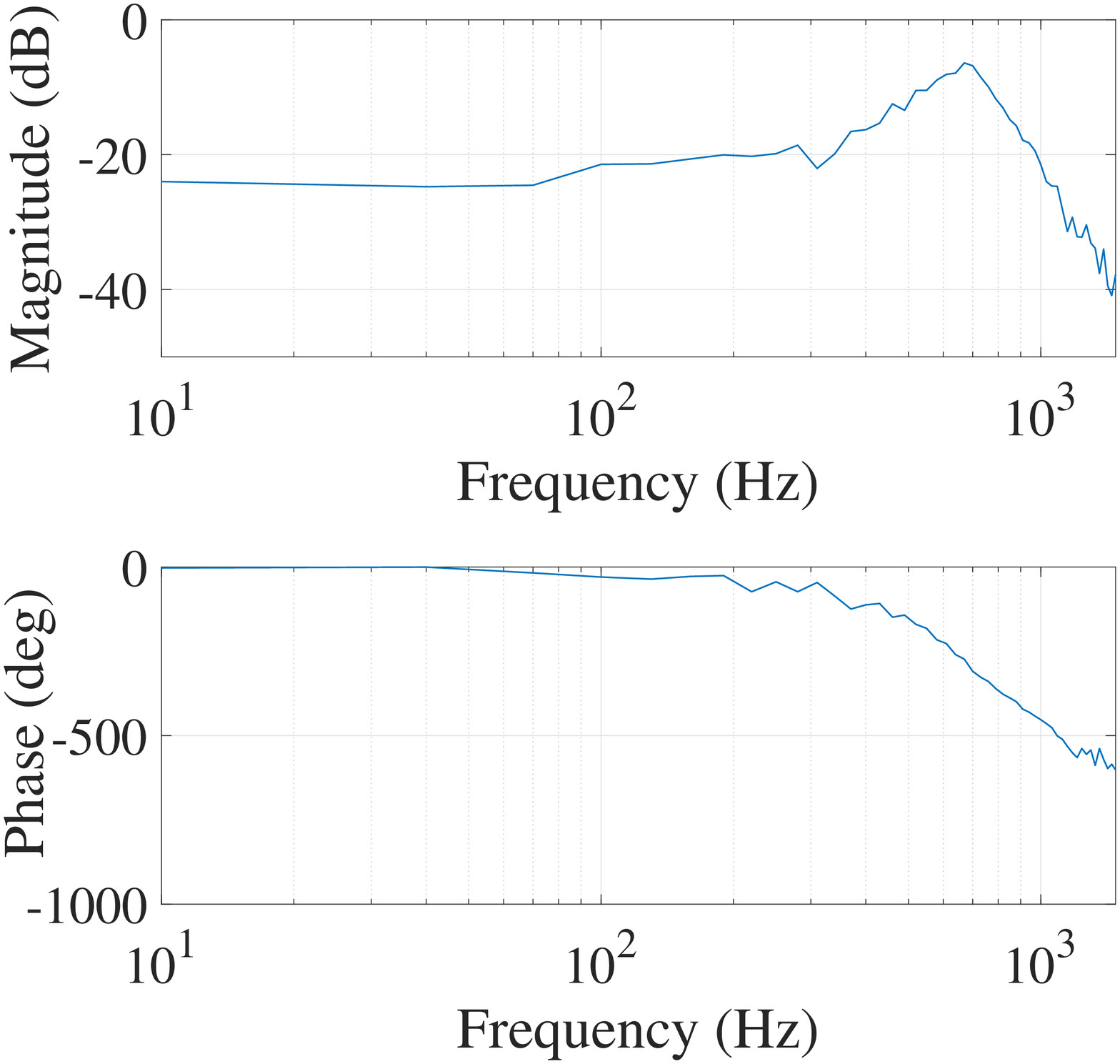}
	\caption{\label{fig:40lpm f resp}50 lpm}
	\end{subfigure}
	\caption{\label{fig:dynamic system ID2}Open-loop Bode plots from ETFE with MC 2, $f_c =$ 2.75 kHz.}
\end{figure}

\begin{figure}
\centering
\captionsetup{justification=centering}
\includegraphics[width=0.7\textwidth]{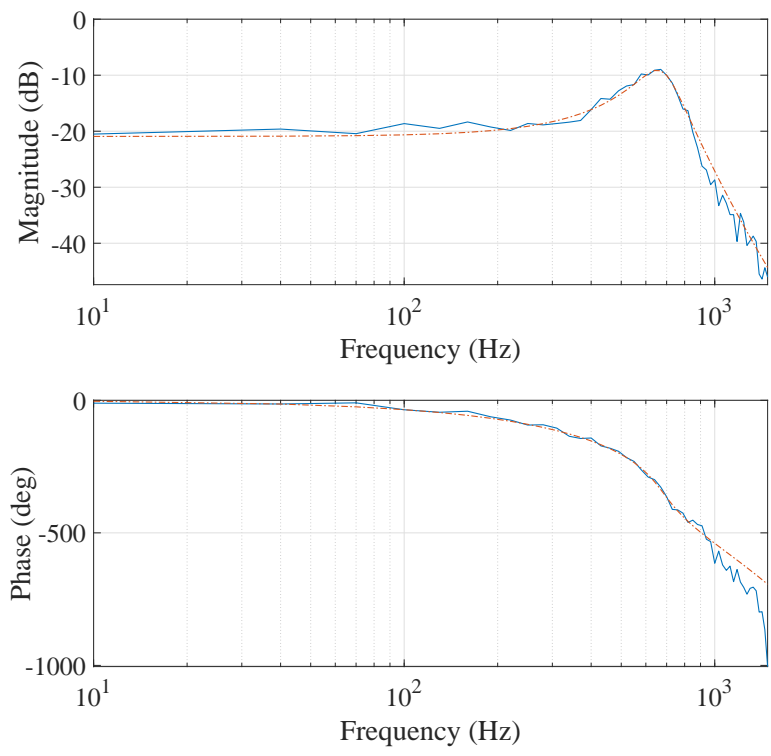}
\caption{\label{fig:OL bode 40 lpm with model TF} Open-loop Bode plot at 40 lpm: ETFE (blue, solid) and fitted transfer function (red, dash-dot).}
\end{figure}

\begin{figure}
\centering
\captionsetup{justification=centering}
\includegraphics[width=\textwidth]{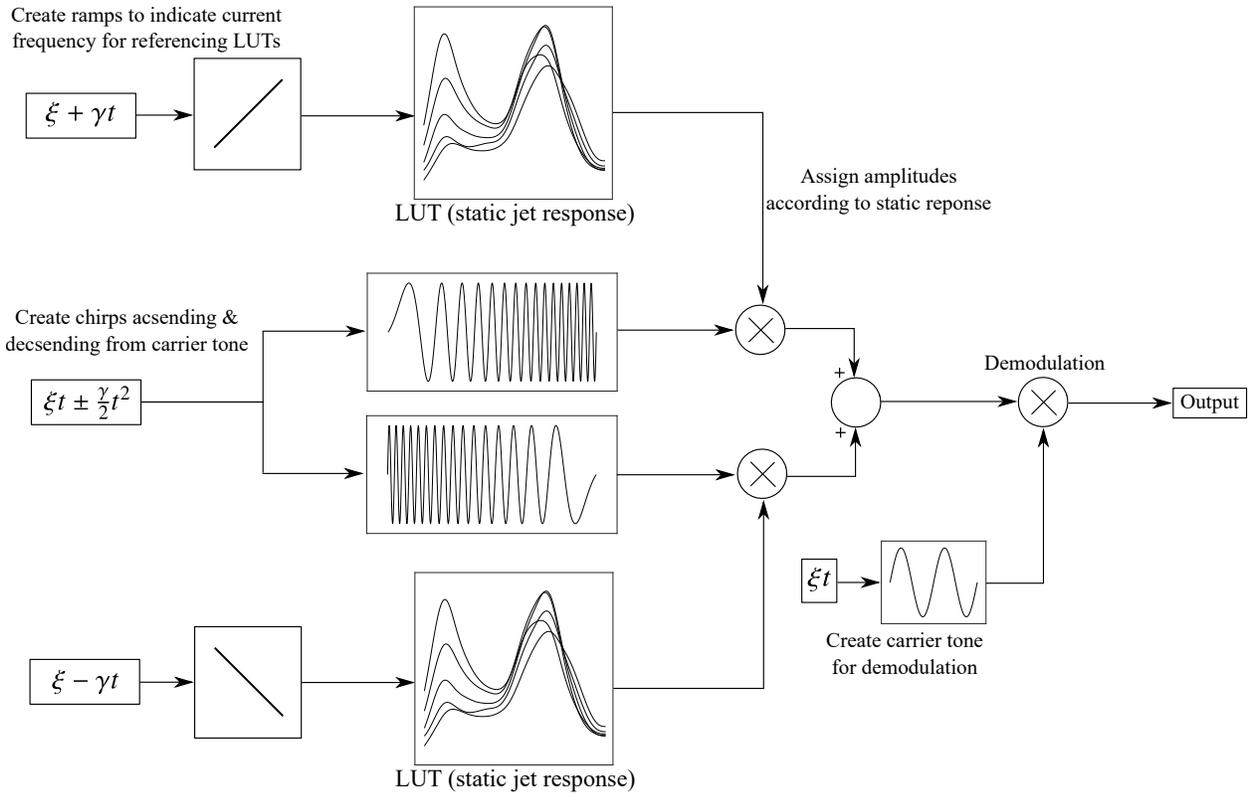}
\caption{\label{fig:simulink diagram for quasi-steady response}Block diagram of simulation used to determine quasi-steady jet response, $\psi(f)$, from static jet deflection response (Fig. \ref{fig:static deflection curves}).}
\end{figure}

\begin{figure}
\centering
\captionsetup{justification=centering}
\includegraphics[width=0.6\textwidth]{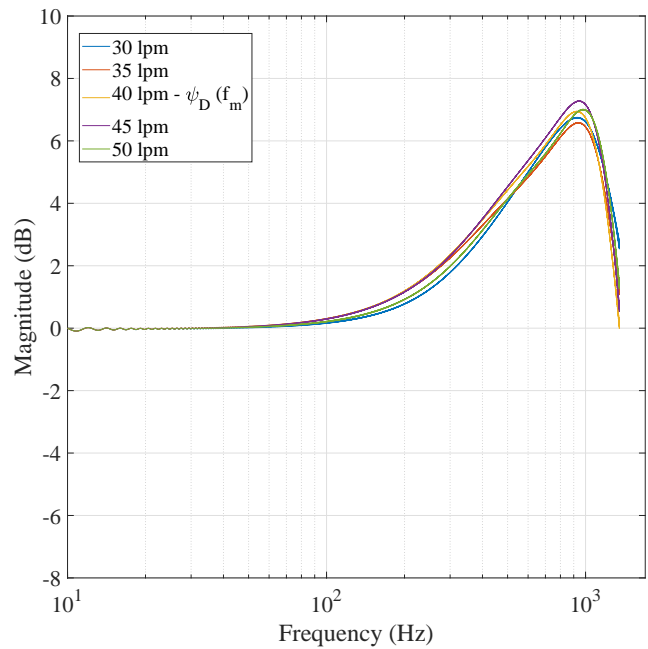}
\caption{\label{fig:quasi-steady jet behaviour}Quasi-steady state jet behaviour, $\psi(f)$, at several flow rates}
\end{figure}

\begin{figure}
\centering
\captionsetup{justification=centering}
\includegraphics[width=0.7\textwidth]{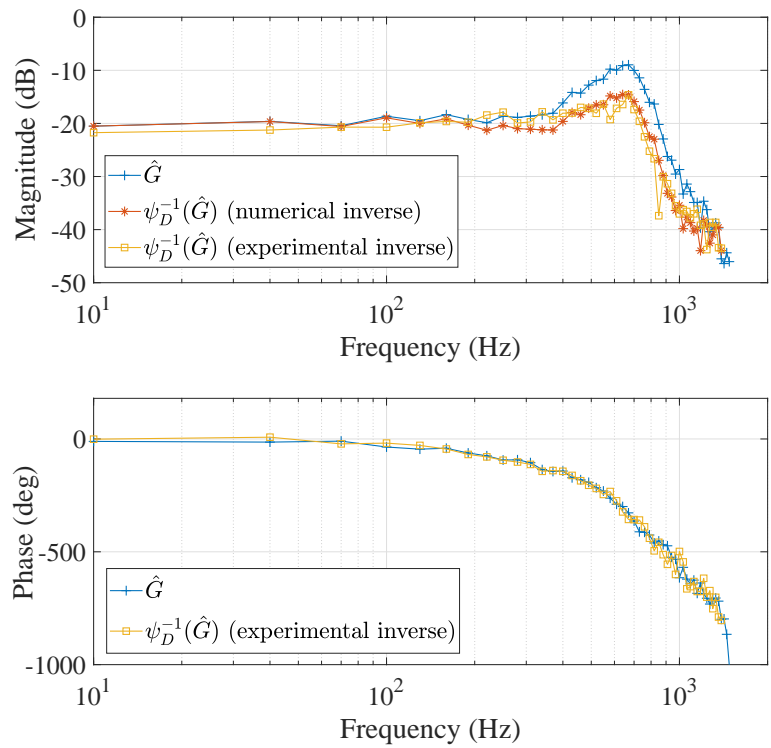}
\caption{\label{fig:ETFE with estimated and measured effect of quasi-steady resp}ETFEs at 40 lpm: original curve, $\hat{G}$ (blue, cross), predicted magnitude response from numerical inversion of quasi-steady jet response, $\psi_D^{-1}\left\{\hat{G}\right\}$ (red, star) and the result of experimentally inverting $\psi_D\left(f_m\right)$ (yellow, square).}
\end{figure}

To identify the jet's dynamic behaviour, we carried out system identifications where the deflection was varied dynamically. The carrier signal at 2.75 kHz, $g_c(t) = sin(2{\pi}f_ct)$ (where $f_c = 2.75$ kHz), is amplitude modulated by another signal, $g_m(t)$. The resulting jet is deflected dynamically and the deflection follows the shape of $g_m(t)$. As an example, if $g_m(t) = Asin(2{\pi}f_mt)$, where $f_m$ is low enough to avoid exciting the jet dynamics, the signal driving the audio amplifier is given by
\begin{equation}
\label{eq:AM signal}
g_{\text{amp}}(t) = g_c(t)g_m(t) = sin(2{\pi}f_ct)F^{-1}\left\{(Asin(2{\pi}f_mt)\,+\,B)\right\},
\end{equation}
where $F^{-1}(x)$ is the inverted system nonlinearity, i.e. the relevant curve in Fig. \ref{fig:NLs inverted LUT data}, and the deflection varies between that which would result from applying the constant excitations $g_{\text{amp}}(t) = F^{-1}\left\{(B-A)\right\}sin(2{\pi}f_ct)$ and $g_{\text{amp}}(t) = F^{-1}\left\{(B+A)\right\}sin(2{\pi}f_ct)$ at a frequency of $f_m$. The system considered is between the modulating signal $g_m(t)$ and the total pressure in the unattached side channel.
\\
\\
The amplitude of the offset of the carrier signal, $B$ in the example above, was chosen to be 0.3 V$_\text{pp}$ because it is in the middle of the range of possible input amplitudes, and the modulation signal amplitude, $A$ above, was also 0.3 V$_{\text{pp}}$. The FPGA produces the signal $g_{\text{amp}}(t) = g_c(t)F^{-1}\left\{g_m(t) + B\right\}$. The signal $g_m(t)$ is a sinusoid which changes in frequency from 10 to 1960 Hz in 30 Hz steps over 200 s, so that
\begin{equation}
\label{eq:chirp signal}
g_m(t) \, = \, A\text{sin}\left(2{\pi}t\left(f_0 + f_1(t)\right)\right),
\end{equation}
where $f_1(t)$ increments by 30 Hz periodically. We chose a stepped sinusoid input rather than a chirp signal because of the high noise levels in the device. A chirp signal spreads the input energy over a broad, continuous range of frequencies, and the jet response at each frequency was found to be dominated by its random fluctuations rather than the response to the input excitation. While the stepped sinusoid does not result in a continuous range of frequencies, the response it produces dominates at a discrete set of frequencies, thus allowing accurate magnitude and phase calculations. This was done at several flow rates around the design case (40 lpm) in order to indicate the sensitivity of the plant to variations in the inlet flow rate, with the relevant nonlinearity compensator, $F^{-1}(x)$, from Fig. \ref{fig:NLs inverted LUT data}, implemented in a LUT.
The empirical transfer function estimate (ETFE) is defined as
\begin{equation}
\hat{G}(\omega) \, = \, \frac{{\Phi}_{yu}(\omega)}{{\Phi}_{uu}(\omega)},
\end{equation}
where ${\Phi}_{yu}(\omega)$ and ${\Phi}_{uu}(\omega)$ are the cross-spectral density of the output and the input and the power spectral density of the input respectively. Note that the input is taken as $u(t) = g_m(t)$. Initially, the ETFE captured at 40 lpm using PT A and measurement connection 1 (MC 1) gave the blue response in Fig \ref{fig:dynamic system ID1}.
It was found that the roll-off in the response, at $\sim$ 150 Hz, was a result of the dynamics of the tube connecting the pressure tapping to the pressure transducer (MC 1), despite the faster response time of the transducer itself (100 $\mu$s). As such, the connection between the pressure tapping and the transducer was redesigned (MC 2), and the experiment was repeated. The corresponding ETFE is shown in red in Fig. \ref{fig:dynamic system ID1}. Additionally, the same system identification was carried out at several flow rates in order to indicate the sensitivity of the plant to this parameter. The ETFEs of these data at each flow rate are shown in Fig. \ref{fig:dynamic system ID2}.
The true jet deflection system roll-off was measured to be slightly higher by repeating the system identification experiments with PT B, which requires no measurement connection and has a higher bandwidth limitation. However, PT B suffers from poor temperature compensation which causes the mean signal to vary, making it unsuitable for use in a tracking problem. Therefore, the small reduction in bandwidth was considered acceptable and PT A was used for the purposes of control. A transfer function model was fitted to $\hat{G}(\omega)$ for the design flow rate, 40 lpm, which is shown in Fig. \ref{fig:OL bode 40 lpm with model TF}.
The transfer function was found using MATLAB's System Identification Toolbox, and is given by
\begin{equation}
\label{eq:open-loop transfer function}
\centering
H_{\text{plant}} \, = \,z^{-36}\frac{4.86{\times}10^{-6}z^{-1}}{1 - 3.88z^{-1} + 5.67z^{-2} - 3.68z^{-3} + 0.899z^{-4}}.
\end{equation}
The estimated input-output delay is 36 time steps, which is 0.72 ms and corresponds to a transport delay in the device. This is the time taken for the jet to travel from the inlet orifice, where upon it is acted by the piezo, to the pitot probe in the unattached side outlet channel. This distance is approximately 50 mm, which gives an time-averaged jet speed of 69 m$\text{s}^{\text{-1}}$ between these points. Based on the orifice mean jet velocity of around 90 m$\text{s}^{\text{-1}}$, this seems to be a reasonable value.
\\
\\
The curves in Fig. \ref{fig:dynamic system ID2} demonstrate a significant variation in the plant dynamics with inlet flow rate away from the 40 lpm design case. In practice, it is possible that the device could be supplied with an unsteady pressure ratio, leading to varying and potentially unknown mass flow rates. The use of closed-loop control makes the system robust to these variations to some degree, which will be tested later.
\\
\\
We now consider the effect of the static deflection curves on the dynamic response. It is well known that the frequency content of an amplitude modulated signal is given by the sum and difference frequencies of the carrier and modulation signals,
\begin{equation}
\label{eq:AM signal breakdown}
sin(2{\pi}f_ct)\left(Asin(2{\pi}f_mt) + B\right) = \frac{A}{2}\left[cos(2{\pi}\left(f_c - f_m\right)t) - cos(2{\pi}\left(f_c + f_m\right)t)\right] + Bsin(2{\pi}f_ct).
\end{equation}
Wiltse and Glezer \cite{wiltse1993manipulation} demonstrated that jets demodulate amplitude modulated acoustic signals and the bulk jet response is seen at the modulation frequency if the amplitude is high enough. This can be expressed mathematically by multiplying the signal by the carrier tone and then low-pass filtering. In practice, the tones on the right-hand side in (\ref{eq:AM signal breakdown}) have different magnitudes from one another according to the shape of the relevant static deflection curve in Fig. \ref{fig:static deflection curves}. The effective magnitude response resulting from this variation alone (i.e. ignoring any jet or sensor dynamics) at a given frequency $f_m$ is therefore a combination of the magnitude of the static deflection curve in Fig. \ref{fig:static deflection curves} at $f_c - f_m$ and $f_c + f_m$, along with the DC offset in jet position in response to the term $Bsin(2{\pi}f_ct)$ in (\ref{eq:AM signal breakdown}). This effective magnitude response, which has no corresponding phase response, is referred to as the quasi-steady jet behaviour. If we fix $f_c = 2750$ Hz, this can be written as a single-input function, which we name $\psi(f)$. The overall system dynamic response to the input (\ref{eq:AM signal breakdown}), given that the amplitude nonlinearity, $F(x)$, has been compensated for, is then given by
\begin{equation}
\label{eq: Overall system dynamic response}
y(t) = \psi(f_m)\,\,|G(2{\pi}f_m)|\,\,sin\left(2{\pi}f_mt + \,\phi\left(G(2{\pi}f_m)\right)\,\right),
\end{equation}
where $G$ is the true jet dynamic transfer function, $|G|$ is its magnitude response and $\phi(G)$ is its phase response. In the frequency domain, without compensating for $\psi(f)$, the  ETFE captured from system identification experiments is
\begin{equation}
\label{eq: Overall system dynamic response frequency domain ETFE}
\hat{G}(2{\pi}f) = \psi(f)G(2{\pi}f).
\end{equation}
To determine the quasi-steady jet behaviour, $\psi(f)$, a simulation was created where two tones, initially at 2750 Hz and out of phase, linearly increased and decreased in frequency respectively at a rate of 28 Hz$\text{s}^{\text{-1}}$ over 50 seconds. This corresponds to two copies of the signal in (\ref{eq:chirp signal for static sys ID}), with $\xi = 2750$ Hz, $\gamma = \pm 28$ Hz$\text{s}^{\text{-1}}$. The magnitude of these tones were assigned by reference to a look-up table, which interpolated the values of the relevant static deflection curve in Fig. \ref{fig:static deflection curves}. The tones were then summed and the signal demodulated by multiplying by the 2750 Hz carrier tone. A block diagram of the simulation is shown in Fig. \ref{fig:simulink diagram for quasi-steady response}.
The power spectral density of the resulting signal, $\psi(f)$, is plotted up to 1350 Hz in Fig. \ref{fig:quasi-steady jet behaviour} at all of the flow rates considered.
\\
\\
The curve for 40 lpm is referred to as $\psi_D(f)$. These responses in Fig. \ref{fig:quasi-steady jet behaviour} justify our choice of carrier tone at 2750 Hz, since the resulting responses are independent of flow rate to within 1 dB, and the roll-off is higher than that of the dynamic response. To consider the effect of these quasi-steady curves on the dynamic response, we repeated the dynamic response identification at the design flow rate (40 lpm) with the amplitude of the modulating tone set to invert the shape of the 40 lpm quasi-steady response curve in Fig. \ref{fig:quasi-steady jet behaviour}. This was implemented in a look-up table ($\psi_D^{-1}\left\{f\right\}$), which was referenced by the frequency of the modulating tone being produced, such that the audio amplifier signal was given by
\begin{equation}
\label{eq:dynamic identification input with compensation}
g_{\text{amp}}(t) = sin(2{\pi}f_ct)F^{-1}\left\{{\psi_D}^{-1}\left\{f_0 + f_1(t)\right\}\text{sin}\left(2{\pi}t\left(f_0 + f_1(t)\right)\right) + B\right\}.
\end{equation}
While it would be more precise to set the amplitudes of the sum and difference tones resulting from the product of the modulating and carrier tones according to the inverse of the 40 lpm static deflection curve in Fig. \ref{fig:static deflection curves}, as in the simulation described above, this was not possible in practice due to the static nonlinearity compensator, $F^{-1}(x)$, in (\ref{eq:dynamic identification input with compensation}), which makes it impossible to evaluate the first term on the right-hand side of (\ref{eq:dynamic identification input with compensation}) analytically.
\\
\\
A dynamic system identification was conducted at 40 lpm using (\ref{eq:dynamic identification input with compensation}), with the carrier offset set to 0.25 $V_{\text{pp}}$ ($B$ in (\ref{eq:dynamic identification input with compensation})). The expected result was predicted by numerically inverting $\psi_D\left(f_m\right)$ and applying it to the original 40 lpm ETFE in Fig. \ref{fig:OL bode 40 lpm with model TF}, i.e. rearranging (\ref{eq: Overall system dynamic response frequency domain ETFE}) to give $G = \psi_D^{-1}\left\{\hat{G}\right\}$. The ETFE resulting from the experiment, the predicted result given by $\psi_D^{-1}\left\{\hat{G}\right\}$, and the original ETFE for 40 lpm from Fig. \ref{fig:OL bode 40 lpm with model TF} are shown in Fig \ref{fig:ETFE with estimated and measured effect of quasi-steady resp}.
There is good agreement between the predicted and measured magnitude responses, and the phase response is the same as the original ETFE data, $\hat{G}$, as expected. This demonstrates that (\ref{eq: Overall system dynamic response frequency domain ETFE}) is a reasonable model for the effect of the quasi-steady jet response. However, it was not possible to implement $\psi_D^{-1}\left\{f_m\right\}$ in practice in a real-time controller because it is defined in the frequency domain and has no corresponding phase response, such that a linear filter could not be used as a model. Additionally, we ignored the 36 $T_\text{s}$ transport delay in the plant model (\ref{eq:open-loop transfer function}) for the purposes of linear control, which is justified in Section \ref{sec:controller design and simulation}.
%
%
\section{Controller Design and Simulation}
\label{sec:controller design and simulation}
\begin{figure}
\centering
\captionsetup{justification=centering}
\includegraphics[width=0.7\textwidth]{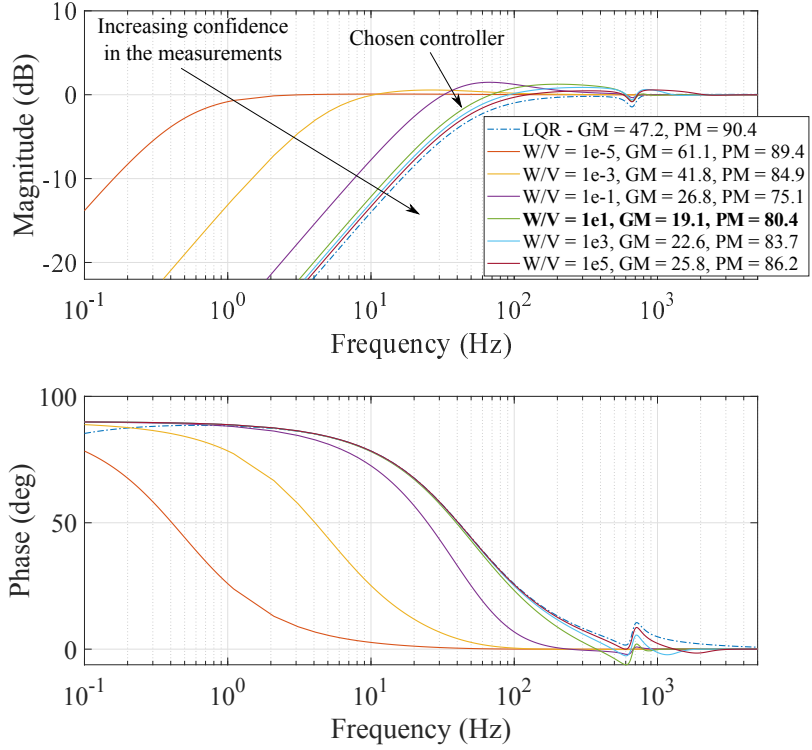}
\caption{\label{fig:sensitivity functions and LTR} LTR procedure demonstrated with sensitivity functions for a variety of process-to-sensor noise ratios. Units for gain and phase margins in legend are dB and degrees respectively. LQR curve is dash-dot blue.}
\end{figure}

\begin{figure}
\centering
\captionsetup{justification=centering}
\includegraphics[width=0.6\textwidth]{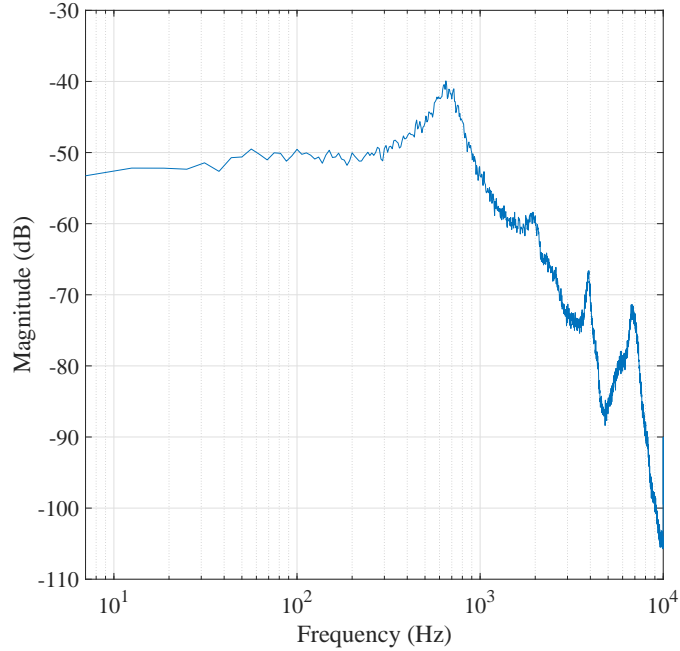}
\caption{\label{fig:noise spectrum}Jet spectrum: PSD of total pressure in unattached side channel at 40 lpm, no excitation.}
\end{figure}

\begin{figure}
\centering
\captionsetup{justification=centering}
\includegraphics[width=0.7\textwidth]{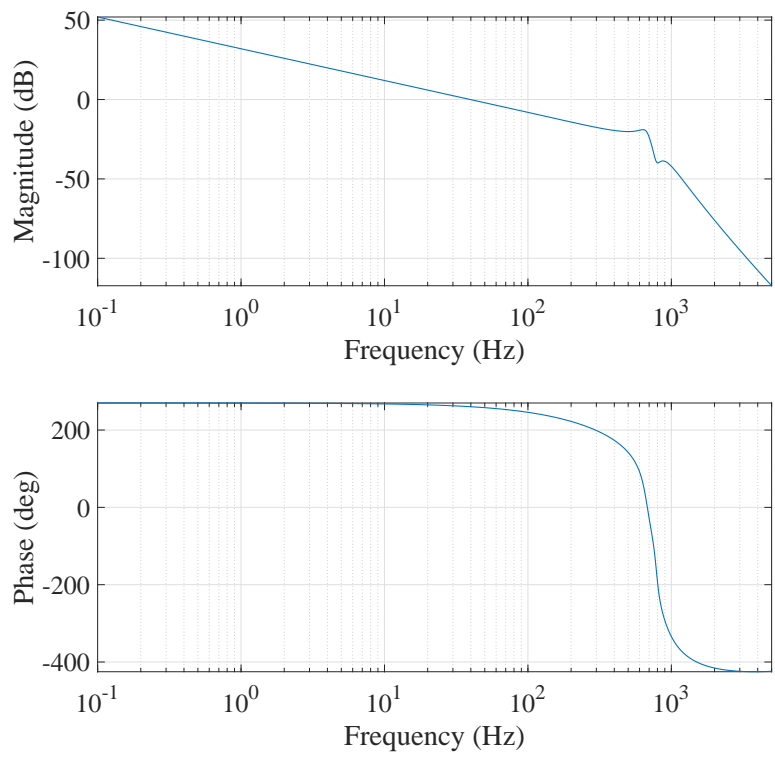}
\caption{\label{fig:LQG TF}LQG transfer function Bode plot}
\end{figure}

We adopted an LQR control strategy, and therefore required an observer to estimate the system states. Converting (\ref{eq:open-loop transfer function}) to a state-space system, 
\begin{align}
\begin{split}
\label{eq: discrete state-space system}
\mathbf{x}(t+T) \, = \, \mathbf{Ax}(t) \, + \, \mathbf{B}u(t)
\\
y(t) \, = \, \mathbf{Cx}(t) + Du(t),
\end{split}
\end{align}
gave the matrices
\begin{align}
\begin{split}
\label{eq: discrete state-space system}
\mathbf{A} = 
\left [ {\begin{array}{cccc}
\begin{aligned}
		3.8&8\quad{\text{-}}2.83\quad\;1.84\quad\;\text{-}0.899
\\		\;2.0&0\quad\quad\;0\quad\quad\;\; 0\quad\quad\;\;\;\; 0
\\		0&\quad\quad1.00\quad\quad0\quad\quad\;\;\;\:\,0
\\		0&\quad\quad\;\;\,0\quad\;\;\,0.500\quad\quad\;0
\end{aligned}
		\end{array}} \right ];\quad
D = 0; \quad\quad\quad
\\
\\
\mathbf{B} =
\left [ {\begin{array}{c}
\begin{aligned}
		0.0&0391 \\
		&0\\
		&0\\
		&0
\end{aligned}
		\end{array}} \right ]; \quad
\mathbf{C} =
\left [ {\begin{array}{ccc}
		0.00297\quad 0 \quad 0\quad 0
		\end{array}} \right ].
\end{split}
\end{align}
The equilibrium of the system is shifted by subtracting the steady state values of the states, input and output once the output has been driven to a reference signal, $r$. These are $\mathbf{x_0}$, $u_0$, and $r$ respectively. New variables are introduced to describe variation from the steady state values, namely $\mathbf{{\delta}x}(t) = \mathbf{x}(t) - \mathbf{x_0}$, ${\delta}u(t) = u(t) - u_0$, and ${\delta}y(t) = y(t) -  r$, so that the system dynamics take the form
\begin{align}
\begin{split}
\label{eq: discrete state-space system steady state shift}
\mathbf{{\delta}x}(t+T) \, &= \, \mathbf{A{\delta}x}(t) \, + \, \mathbf{B}{\delta}u(t)
\\
{\delta}y(t) \, &= \, \mathbf{C{\delta}x}(t).
\end{split}
\end{align}
To eliminate steady state error, the plant is augmented with an integrator state defined by
\begin{align}
\begin{split}
\label{eq: integrator state}
z(t+T) \, = \, z(t) \,+ \, {\delta}y(t),
\end{split}
\end{align}
such that when process and sensor noise is added, the dynamics become
\begin{align}
\begin{split}
\label{eq: discrete state-space system augmentation}
\left [ {\begin{array}{c}
\begin{aligned}
		\mathbf{{\delta}x}(t+T)
\\		z(t+T)
\end{aligned}
		\end{array}} \right ] \, = \,
\left [ {\begin{array}{cc}
\begin{aligned}
		\mathbf{A} \quad\quad 0
\\		\mathbf{C} \quad\quad 1
\end{aligned}
		\end{array}} \right ]
\left [ {\begin{array}{c}
\begin{aligned}
		\mathbf{{\delta}x}(t)
\\		z(t)
\end{aligned}
		\end{array}} \right ] \, + \,
\left [ {\begin{array}{c}
\begin{aligned}
		\mathbf{B}
\\		0
\end{aligned}
		\end{array}} \right ]
{\delta}u(t) \, + \,
\left [ {\begin{array}{c}
\begin{aligned}
		\mathbf{F}
\\		0
\end{aligned}
		\end{array}} \right ]
w(t)
\\
\left [ {\begin{array}{c}
\begin{aligned}
		{\delta}y'(t)
\\		z'(t)
\end{aligned}
		\end{array}} \right ] \, = \,
\left [ {\begin{array}{cc}
\begin{aligned}
		\mathbf{C}& \quad\quad 0
\\		0& \quad\quad 1
\end{aligned}
		\end{array}} \right ]
\left [ {\begin{array}{c}
\begin{aligned}
		\mathbf{{\delta}x}(t)
\\		z(t)
\end{aligned}
		\end{array}} \right ]  \, + \,
\mathbf{v}(t), \quad\quad\quad\quad\quad\quad\,\,\,
\end{split}
\end{align}
where ${\delta}y'(t)$ and $z'(t)$ are the noisy measurements of ${\delta}y(t)$ and $z(t)$. The regulator minimised the cost function given by
\begin{equation}
\label{eq:lqr cost function}
J \, = \,\sum_{i=0}^{i=\infty}\left({\delta}\mathbf{x}_i^\text{T}\mathbf{Q}{\delta}\mathbf{x}_i \, + \, R{\delta}u_i^2\right),
\end{equation}
where ${\delta}\mathbf{x}_i = {\delta}\mathbf{x}(t+iT)$ and ${\delta}u_i = {\delta}u(t + iT)$. The tuned LQR cost parameters are 
\begin{align}
\begin{split}
\label{eq:LQR cost matrices}
\mathbf{Q} =
\left [{\begin{array}{ccccc}
\begin{aligned}
&1 \quad \;0 \quad \;0 \;\quad0\;\quad\;0\\
&0 \quad \;0 \quad \;0 \;\quad0\;\quad\;0\\
&0 \quad \;0 \quad \;0 \;\quad0\;\quad\;0\\
&0 \quad \;0 \quad \;0 \;\quad0\;\quad\;0\\
&0 \quad \;0 \quad \;0 \;\quad0\;\quad10^3
\end{aligned}
	\end{array}} \right], \quad
 R = 10^6,
\end{split}
\end{align}
where $\mathbf{Q}$ is the state-error cost matrix and $R$ is the cost of control actions. $\mathbf{Q}$ was chosen as per standard output-weighted LQR, i.e. $\mathbf{C}^\text{T}\mathbf{C}$, and the integrator cost was tuned manually. The states were not measured directly so a Kalman filter was used for state estimation. The observer dynamics take the form
\begin{equation}
\label{eq:estimation dynamics}
\begin{aligned}
	\left [{\begin{array}{c}
	{\delta}\mathbf{\hat{x}}\left(t+T\right)\\
	\hat{z}(t+T)
	\end{array}} \right] \, = \,
	\left [{\begin{array}{cc}
	\mathbf{A} & 0\\
	\mathbf{C} & 1
	\end{array}} \right]
	\left [{\begin{array}{c}
	{\delta}\mathbf{\hat{x}}(t)\\
	\hat{z}(t)
	\end{array}} \right] \, + \,
	\left [{\begin{array}{c}
	\mathbf{B}\\
	0
	\end{array}} \right]\,{\delta}u(t) \, + \,
	\mathbf{K}_f
	\left [{\begin{array}{c}
	{\delta}y'(t) - {\delta}\hat{y}(t)\\
	z'(t) - \hat{z}(t)
	\end{array}} \right]
\end{aligned}
\end{equation}
\begin{equation}
\label{eq:output equation observer}
\begin{aligned}
{\delta}\hat{y}(t) \, = \,
	\left [{\begin{array}{cc}
	\mathbf{C} & 0
	\end{array}} \right]
	\left [{\begin{array}{c}
	{\delta}\mathbf{\hat{x}}(t)\\
	\hat{z}(t)
	\end{array}} \right],
\end{aligned}
\end{equation}
where $\hat{{\delta}\mathbf{x}}$, $\hat{{\delta}y}$ and $\hat{z}$ are the estimates of the state deviations, the output error and the integrator respectively. The augmented system matrices are substituted as follows
\begin{equation}
\label{eq:augmented system substitution}
\begin{aligned}
\mathbf{A}_\text{aug} \, = \,
\left [{\begin{array}{cc}
	\mathbf{A} & 0\\
	\mathbf{C} & 1
	\end{array}} \right], \quad
\mathbf{B}_\text{aug} \, = \,
\left [{\begin{array}{c}
	\mathbf{B}\\
	0
	\end{array}} \right], \quad
\mathbf{C}_\text{aug} \, = \,
\left [{\begin{array}{cc}
	\mathbf{C} & 0\\
	\mathbf{0} & 1
	\end{array}} \right], \\
\mathbf{F}_\text{aug} \, = \,
\left [{\begin{array}{c}
	\mathbf{F}\\
	0
	\end{array}} \right],\quad
\mathbf{x}_\text{aug} \, = \,
\left [{\begin{array}{c}
	{\delta}\mathbf{x}\\
	z
	\end{array}} \right], \quad
\mathbf{e}_\text{aug} \, = \,
\left [{\begin{array}{c}
	{\delta}\mathbf{x} - {\delta}\mathbf{\hat{x}}\\
	z - \hat{z}
	\end{array}} \right],
\end{aligned}
\end{equation}
leading to closed-loop dynamics given by
\begin{equation}
\label{eq:LQG combined dynamics}
\begin{aligned}
\left [{\begin{array}{c}
	\mathbf{x}_\text{aug}(t+T)\\
	\mathbf{e}_\text{aug}(t+T)
	\end{array}} \right] \, = \,
\left [{\begin{array}{cc}
	\mathbf{A}_\text{aug} - \mathbf{B}_\text{aug}\mathbf{K}_\text{LQR} & \mathbf{B}_\text{aug}\mathbf{K}_\text{LQR}\\
	\mathbf{0} & \mathbf{A}_\text{aug} - \mathbf{K}_f\mathbf{C}_\text{aug}
	\end{array}} \right]
\left [{\begin{array}{c}
	\mathbf{x}_\text{aug}(t)\\
	\mathbf{e}_\text{aug}(t)
	\end{array}} \right] \, + \,
\left [{\begin{array}{cc}
	\mathbf{F}_\text{aug} & \mathbf{0}\\
	\mathbf{F}_\text{aug} & -\mathbf{K}_f
	\end{array}} \right]
\left [{\begin{array}{c}
	w(t)\\
	\mathbf{v}(t)
	\end{array}} \right],
\end{aligned}
\end{equation}
where $\mathbf{e}_\text{aug}$ is the state estimation error.
\\
\\
There is no automatic guarantee of robustness despite the individual gain and phase margin guarantees of the Kalman filter and LQR separately in continuous-time \cite{doyle1978guaranteed}. Furthermore, in the practical application considered here, discrete-time LQR has typically inferior stability margin properties \cite{shaked1986guaranteed}. It is well known that the loop-transfer recovery (LTR) methodology presented by Doyle and Stein in \cite{doyle1979robustness} for continuous-time systems allows the robustness properties of the full state feedback case to be recovered. It is also well understood that there lies a trade off between recovering these robustness margins and the system noise performance in continuous-time. In discrete-time it may not be possible to recover fully the LQR loop transfer properties \cite{maciejowski1985asymptotic}.
The value of the sensitivity transfer function across the frequency space,
\begin{equation}
S(z) = \frac{1}{1 + C(z)H_{\text{plant}}(z)},
\end{equation}
determines a controller's ability to reject disturbances, follow the reference signal as well its noise performance \cite{bode1945network}.
\\
\\
The sensitivity transfer functions of the system with several process-to-sensor noise ratios are shown in Fig. \ref{fig:sensitivity functions and LTR} to demonstrate the LTR procedure. The integrator state measurement noise variance is fixed at 100$\sigma_{y'}^2$, where $\sigma_{y'}^2$ is the sensor noise variance for the measurement ${\delta}y'(t)$.
Fig. \ref{fig:sensitivity functions and LTR} shows that the LQG curves approach the LQR curve as the process-to-sensor noise ratio is increased. However, practical controller design is not as simple as choosing the system found by allowing $\frac{W}{V}\to\infty$, which would likely suffer from poor noise performance.
In order to select a value of  $\frac{W}{V}$, the system noise spectrum must be considered. A time series was captured by sampling the unattached side channel pressure tapping whilst the jet was unexcited. The turbulence is technically an output disturbance but cannot be rejected by control action - the acoustic signals only control the jet position. Therefore, the turbulence can be thought of as sensor noise for control purposes. The power spectral density (PSD) of the time series is shown in Fig. \ref{fig:noise spectrum}.
As can be seen in Fig. \ref{fig:noise spectrum}, the roll-off is caused in part by the measurement system, as described in section \ref{sec:dynamic ID}, and the frequency content is in the same bandwidth as the system. This means that choosing a larger $\frac{W}{V}$ causes the controller not only to reject disturbances that can be controlled (e.g. fluctuations in the inlet mass flow)  and follow the reference more effectively, but also to react to the jet turbulence. Therefore a compromise must be made to give both acceptable disturbance rejection and noise performance. The value of $\frac{W}{V}$ chosen is 1e1, with a gain margin of 19.1 dB and a phase margin of 80.4$^\text{o}$.
The tuned LQR costs and Kalman filter variances that give these results are
\begin{align}
\begin{split}
\label{eq:kalman filter noise variances}
W = 1, \quad\;\mathbf{V} =
\left [{\begin{array}{cc}
\begin{aligned}
&1\text{e-}1 \quad 0 \; \;\\
&\;\;0 \quad 1\text{e}1
\end{aligned}
\end{array}} \right]
\end{split}
\end{align}
where $W$ and $\mathbf{V}$ are the process and sensor noise variance and covariance respectively. Note that the 5$^\text{th}$ state in the system is the integrator state, and that the 2$^\text{nd}$ dimension in the covariance matrix $V$ is the variance of the integrator state measurement. The LQG controller magnitude response is shown in Fig. \ref{fig:LQG TF}.
The cross-over frequency for this controller is around 50 Hz, while the transport delay (0.72 ms) is a period corresponding to $\sim$ 1400 Hz. Therefore, since the closed-loop bandwidth is two orders of magnitude below this, our earlier decision not to include the input-output delay in the controller design process is justified.
\section{Implementation and Results}
\label{sec:implementation and results}
\begin{figure*}
\centering
\captionsetup{justification=centering}
\includegraphics[width=\textwidth]{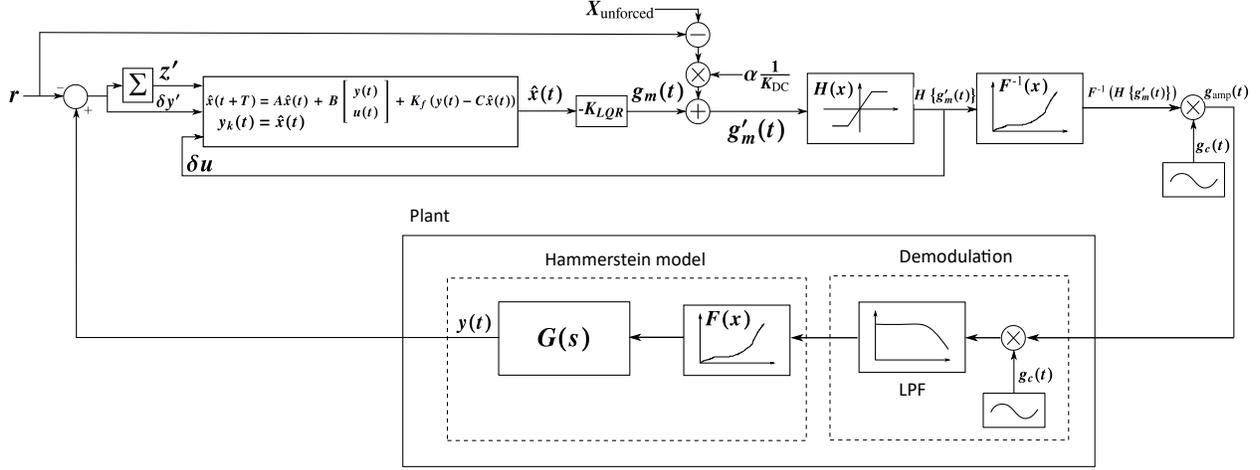}
\caption{\label{fig:closed-loop block diagram} Closed-loop system block diagram}
\end{figure*}

\begin{figure}
\centering
\captionsetup{justification=centering}
\includegraphics[width=0.8\textwidth]{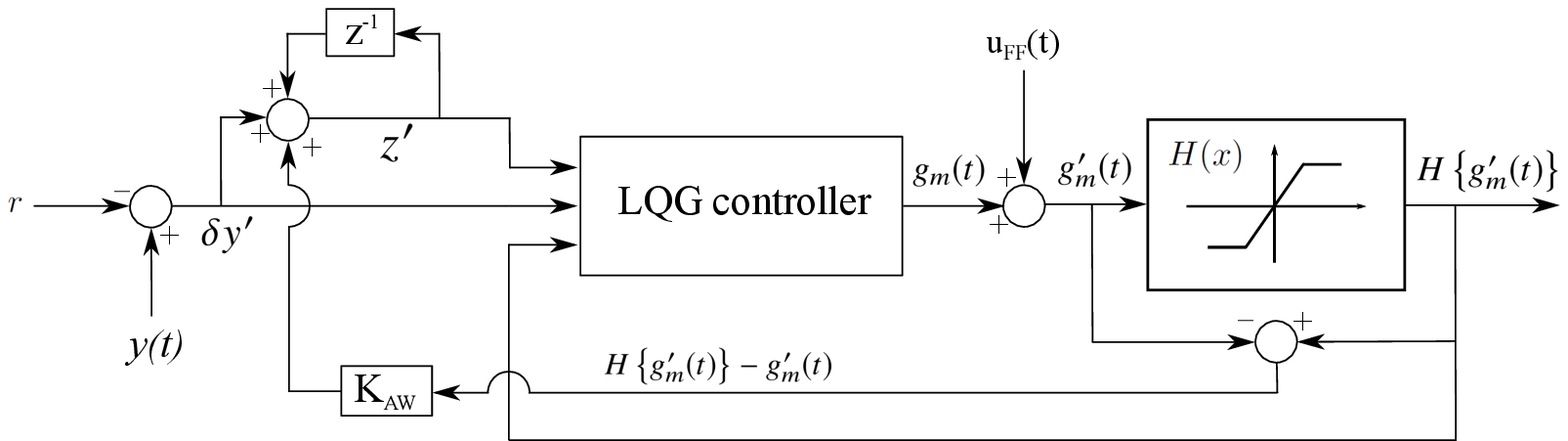}
\caption{\label{fig:anti wind-up scheme}Anti wind-up scheme}
\end{figure}

The LQG controller simulated in Sec. \ref{sec:controller design and simulation} was implemented in the FPGA and the system block diagram is shown in Fig. \ref{fig:closed-loop block diagram}.
As shown in the diagram, a feed-forward term was added to help increase the speed of the initial rise of the step response, giving the control law
\begin{equation}
\label{eq:lqr control law}
u(t) \, = \, {\minus}\mathbf{K}_\text{LQR}
\left [ {\begin{array}{c}
\begin{aligned}
		\hat{{\delta}\mathbf{x}}(t)
\\		\hat{z}(t)
\end{aligned}
		\end{array}} \right ] \, + \, u_\text{FF}(t),
\end{equation}
where $u_\text{FF}(t)$ is the feed-forward term. The DC level measured by the pressure transducer due to the unforced response is $X_\text{unforced}$, $K_\text{DC}$ is the DC gain of the model, and $\alpha$ is a parameter used to vary the contribution from the feed-forward term when tuning the step response. The implementation of the control law is therefore
\begin{equation}
\label{eq:implementation of control law}
g_m'(t) = g_m(t) +  {\alpha}\frac{\left(r - X_\text{unforced}\right)}{K_\text{DC}}.
\end{equation}
It can be seen by comparison of (\ref{eq:lqr control law}) and (\ref{eq:implementation of control law}) that $g_m(t)$ is the feedback term and $u_\text{FF}(t) = {\alpha}\frac{\left(r - X_\text{unforced}\right)}{K_\text{DC}}$. Fig. \ref{fig:closed-loop block diagram} also shows that $g_m'(t)$ was limited to between 0 and 0.8 V, denoted by $H\left\{g_m'(t)\right\}$. This was necessary because the linear model fails to predict the plant behaviour outside this region. If $F^{-1}\left(g_m(t)\right) > $ 0.8 V, the deflection does not increase because of the lack of strict monotonicity of the system characterisation in this range (see Sec. \ref{sec:linearity}). The model indicates that values of $F^{-1}\left(g_m(t)\right) < $ 0 V will cause the jet to deflect away from the centre of the device, which is not the case; the deflection is caused, in steady state, by $g_c(t)$, whose phase does not affect the direction of deflection. The audio amplifier signal is therefore $g_\text{amp}(t) = F^{-1}\left(H\left\{g_m(t)\right\}\right)g_c(t)$.
\\
\\
The combination of an integrator state and input saturation limits necessitates an anti wind-up scheme to avoid an unstable integrator state. The details of this scheme are shown in Fig. \ref{fig:anti wind-up scheme}, where the value of the anti wind-up gain, $K_\text{AW}$, was set by manual tuning, and took a value of $1\text{e-}5$.
This was not included in the overall closed-loop block diagram in Fig. \ref{fig:closed-loop block diagram} for brevity.
%
\subsection{Step response}
\label{sec:step response}
\begin{figure}
\centering
\captionsetup{justification=centering}
	\begin{subfigure}[t]{0.48\textwidth}
	\captionsetup{justification=centering}
	\includegraphics[width=\textwidth]{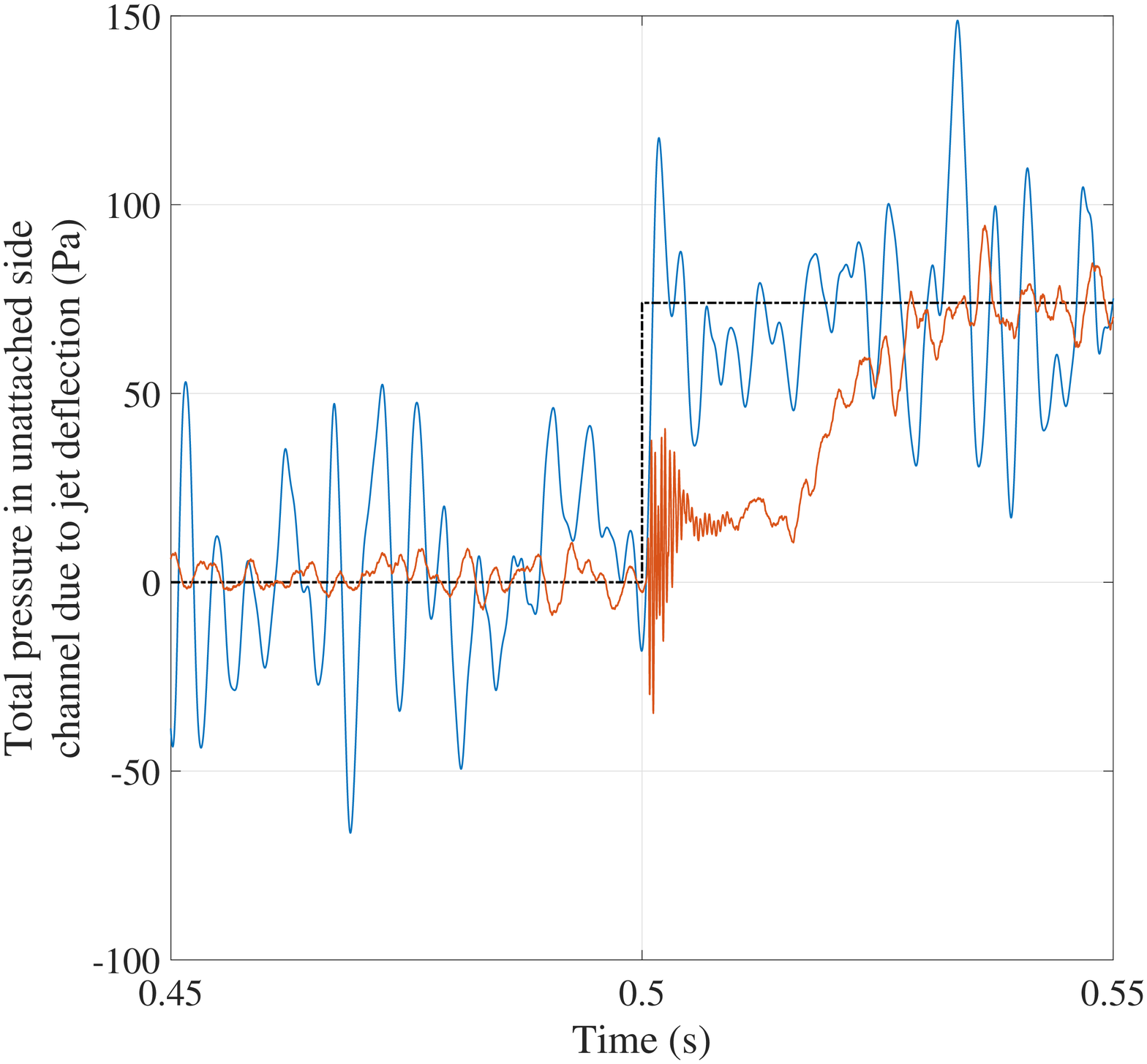}
	\caption{\label{fig:experimental step responses} Ensemble averaged \& filtered step responses (n = 50): closed-loop at 40 lpm (blue) and open-loop full switch at 15 lpm (red). Reference (dash-dot black).}
	\end{subfigure}
	\begin{subfigure}[t]{0.48\textwidth}
	\captionsetup{justification=centering}
	\includegraphics[width=\textwidth]{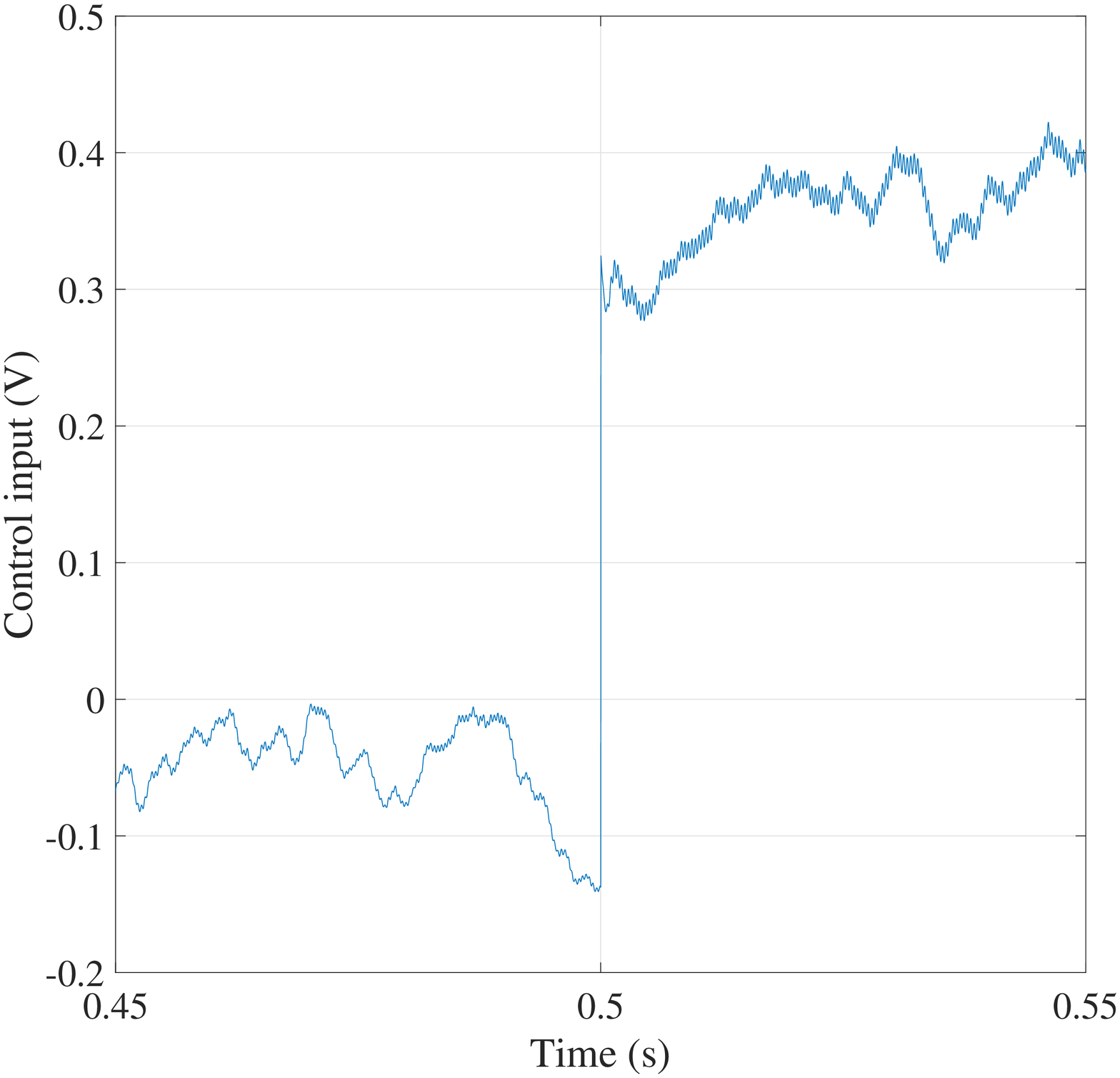}
	\caption{\label{fig:control signal}Control signal (ensemble averaged, n = 50), $g_m'(t)$ in Fig. \ref{fig:closed-loop block diagram}.}
		\end{subfigure}
\caption{System step response: Open- and closed-loop system responses and closed-loop control input.}
\end{figure}

The ensemble averaged (n = 50) response of this system to a reference step from 0 to 74 Pa deflection in the unattached side channel was recorded in closed-loop, with the inlet flow at 40 lpm. As a comparison, an ensemble averaged (n = 50) full switch was recorded at a lower flow rate, such that the pressure in the initially unattached side channel was the same as the deflection in the closed-loop case once the jet had switched, i.e. 74 Pa. To achieve this, the flow rate was set to 15 lpm, and the amplitude of the piezo tone was set to 0.8 V.
The ensemble-averaged time series were then filtered with a notch filter at 2.75 kHz to reduce the direct coupling between the piezo tone and the pressure transducer. A broad notch was also applied at 650 Hz to reduce the jet background noise in order to judge when the deflection first reached its steady state value. The resulting step responses are shown in Fig. \ref{fig:experimental step responses}, while Fig. \ref{fig:control signal} shows an ensemble averaged control signal (n = 50), $g_m'(t)$ (see Fig. \ref{fig:closed-loop block diagram}). In terms of rise time, the closed-loop response at 1.1 ms is thirty times faster than the open-loop full switch at 34 ms. However, this improvement comes from the faster jet dynamics at higher velocities and from using a feed-forward term in the control law rather than the feedback term. Nevertheless, using feed-forward control alone does not compensate for model uncertainty or disturbances such as mass flow variations.
\subsection{Disturbance rejection - mass flow variation}
\label{sec:Disturbance rejection - mass flow variation}
\begin{figure*}
\centering
\captionsetup{justification=centering}
\includegraphics[width=0.6\textwidth]{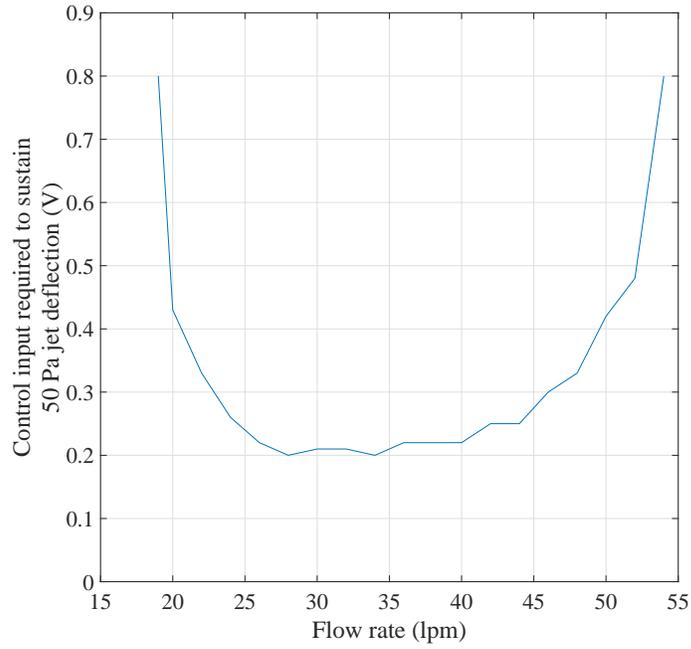}
\caption{\label{fig:flow rate disturb rejection} Mean control input required to maintain 50 Pa jet deflection over several flow rates.}
\end{figure*}

The controller's ability to reject disturbances was tested in two ways. First the inlet mass flow was varied. This was done in the steady state since the response time of the FMA-2612A was insufficient to test the closed-loop bandwidth of the controller. The flow rate was set to values between 19 and 54 lpm, which was the range of flow rates that the controller was able to reject whilst tracking a reference of 50 Pa deflection from the natural bias state. The mean control inputs required to maintain these deflections are shown in Fig. \ref{fig:flow rate disturb rejection}.
The figure shows how the DC gain of the jet deflection system varies over different flow rates, with a broad maximum DC gain at the minimum of the curve, i.e. between 28 and 35 lpm.
\subsection{Disturbance rejection - input disturbance}
\label{sec:Disturbance rejection - input disturbance}
\begin{figure}
\centering
\captionsetup{justification=centering}
\includegraphics[width=0.6\textwidth]{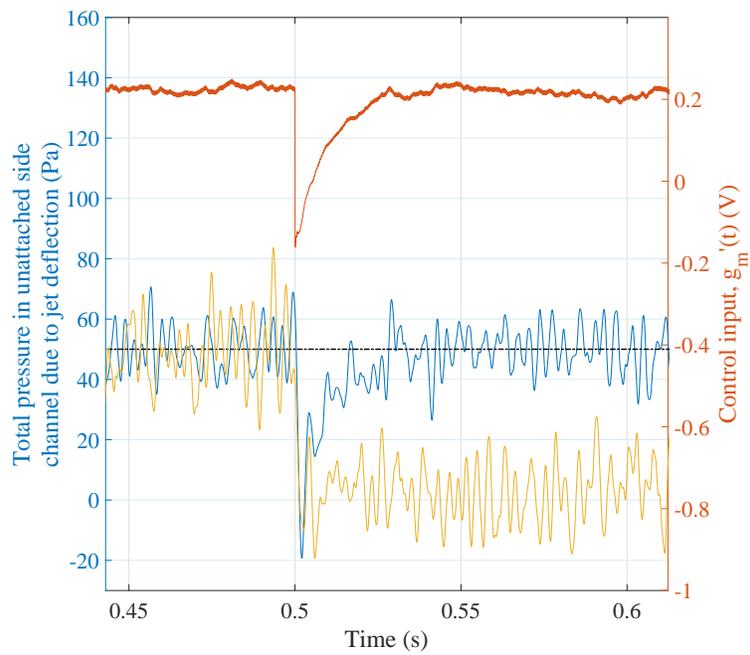}
\caption{\label{fig:disturb rejection input dist} Controller response to input disturbances. Pressure in unattached side channel due to deflection: closed-loop response (blue) and open-loop response (yellow). Control signal ($g_m'(t)$) showing step change in $u_\text{FF}(t)$ at $t=0.5$s (red). Signal processing scheme described in text.}
\end{figure}

\begin{figure}
\centering
\captionsetup{justification=centering}
\includegraphics[width=0.6\textwidth]{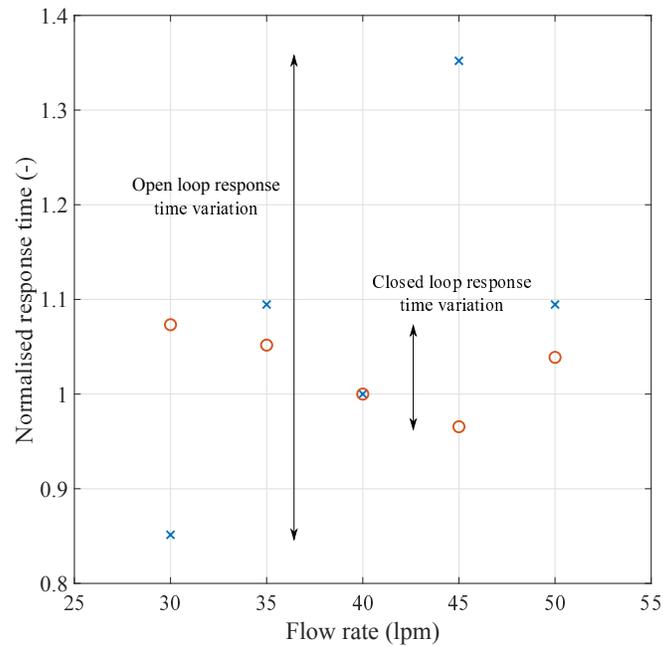}
\caption{\label{fig:response time variation OL CL disturb rej} Controller response time to input disturbances relative to reference case (40 lpm): open-loop (blue, cross) and closed-loop (red, circle).}
\end{figure}

The response of the controller to input disturbances was tested at the design flow rate (40 lpm) by removing the feed-forward term whilst the controller was tracking a constant reference deflection of 50 Pa. The pressure in the unattached side channel due to the deflection and the control input signal are shown in Fig. \ref{fig:disturb rejection input dist}.
The signal processing for these data were as follows: as for the step responses in Section \ref{sec:step response} concerning the step response, a 2.75 kHz high-Q factor notch as well as a broad notch at 650 Hz were applied to ensemble averaged time series (n = 50). This was repeated 6 times and the resulting signals were also ensemble averaged. This additional ensemble averaging step was required due to the higher variance of the open-loop response times (relative to the mean). The input signal was ensemble averaged without any filtering (n = 350). Fig. \ref{fig:disturb rejection input dist} indicates that the closed-loop response time to disturbances is around 28 ms, whereas the case with no feedback term has a faster response time (1.4 ms) but is unable to reject the input disturbance (as expected). It is interesting to consider the response times of the open- and closed-loop cases over several flow rates in order to evaluate the robustness of the input disturbance rejection properties to varying operating conditions. To see this, the same input disturbance experiment described above was carried out at several mass flow rates for both the open- and closed-loop cases. The response time was recorded for each case. For the closed-loop cases, this means the time to return to the reference of 50 Pa deflection, whereas for the open-loop cases, it is the time to reach the natural, undeflected jet position at each flow rate (i.e. 0 Pa in the Fig. \ref{fig:disturb rejection input dist}). Each response time was normalised relative to the 40 lpm cases for open- and closed-loop respectively. These normalised times are shown in Fig. \ref{fig:response time variation OL CL disturb rej}. 
The figure shows that response times vary significantly more in the open-oop case. In the closed-loop case the times vary by 7$\%$, whereas the open-loop times vary by 35$\%$. The ability not only to reject disturbances but also to reject them with a relatively consistent response over a wide range of operating conditions highlights the benefits of using feedback.
\subsection{Performance summary of device control modes}
\label{sec:Performance summary of device control modes}
The response times of the various device control modes are in Table \ref{tab: control mode response time}.
\begin{table}[h]
\centering
\captionsetup{justification=centering}
\begin{tabular}{|c|c|c|c|}
\hline
\textbf{Control mode}                                                                           & \textbf{\begin{tabular}[c]{@{}c@{}}Response time to\\ reference signal (ms)\end{tabular}} & \textbf{\begin{tabular}[c]{@{}c@{}}Response time to input\\ disturbances (ms)\end{tabular}} & \textbf{\begin{tabular}[c]{@{}c@{}}Variation in response time\\ to input disturbances (-)\end{tabular}} \\ \hline
\begin{tabular}[c]{@{}c@{}}Open-loop controller at 15 lpm\\ (full switch)\end{tabular}          & 34                                                                                        & $\infty$                                                                      & 35\%                                                                                                    \\ \hline
\begin{tabular}[c]{@{}c@{}}Closed-loop controller at 40 lpm\\ (partial deflection)\end{tabular} & 1.1                                                                                       & 28                                                                                         & 7\%                                                                                                     \\ \hline
\end{tabular}
\caption{\label{tab: control mode response time}Response times for device control modes}
\end{table}
We draw a comparison between the closed-loop response time to input disturbances, which is effectively the feedback term response time and is 28 ms, and the open-loop response time for the full switch at 15 lpm, i.e. the red curve in Fig. \ref{fig:experimental step responses}, which is 34 ms. With the feedback term alone, we have recovered and improved upon this response time, with the added benefits of robustness to mass flow fluctuations and input disturbances. In practice, our controller also includes the feed-forward term (closed-loop response time to reference signal in Table \ref{tab: control mode response time}), which significantly reduces the time taken to respond to changes in the reference signal (1.1 ms) because of the faster jet dynamics at higher flow rates, while maintaining the robustness to mass flow and input disturbances due to the feedback term.
\section{Conclusions and Future Work}
A novel, bistable fluidic amplifier was controlled by a piezoelectric buzzer. The degree of deflection of the jet was determined by making measurements with a pressure transducer connected to a total pressure tapping in the unattached side channel of the device. The sensitivity of the jet deflection in steady state to piezo tone frequencies was assessed, and the deflection was varied to identify the dynamic response. A discrete linear model was developed from these data, and an LQG regulator was designed to drive the jet deflection to follow a reference. This controller was implemented on an FPGA and the resulting closed-loop controller brought a 90 ms$^{\text{-}1}$ jet under control. The disturbance rejection performance of the controller was evaluated by varying the flow conditions and adding input disturbances.
\\
\\
The flow rates used in the present work were limited by the bandwidth of the piezo transducers used (the carrier tone frequency used was far below the jet column or preferred shear layer mode subharmonics). However, it is shown in \cite{mair2017switching} that ultrasonic piezo actuators can be used to switch the jet at flow rates with inlet-to-outlet pressure ratios up to 1.5. Therefore, in future work these actuators could be used in closed-loop control of higher speed jets.
Current work by the authors involves physical modelling of the amplifier in order to explain frequency responses of the jet deflection measured for each flow rate in Fig. \ref{fig:dynamic system ID2}.
%
\section*{Funding Sources}
\label{sec:funding sources}
The authors gratefully acknowledge the support of this work by Rolls-Royce and the EPSRC program
for active control of fluid flows in gas turbines (EP/L015196/1)

\section*{Acknowledgements}
We would like to thank Michael Mair for the fluidic device design used in this paper.
\singlespacing
\bibliography{references}
\end{document}